\documentclass{ws-ijmpb}

\usepackage{emlines}
\usepackage{graphicx}

\begin{document}

\markboth{
Yu.~Holovatch,
 V.~Blavats'ka,  M.~Dudka,
C.~von~Ferber,  R.~Folk, and
T.~Yavors'kii
}
{
Weak quenched disorder and criticality:
resummation of asymptotic(?) series
}

\catchline{}{}{}

\title{
Weak quenched disorder and criticality:
resummation of asymptotic(?) series
}

\author{\footnotesize Yu. Holovatch}

\address{
Institute for Condensed Matter Physics, National Academy of Sciences
of Ukraine,\\  Lviv, UA--79011, Ukraine\\ Ivan
Franko National University of Lviv,\\ Lviv, UA--79005,Ukraine
}

\author{
V. Blavats'ka and  M. Dudka
}

\address{
Institute for Condensed Matter Physics, National Academy of Sciences
 of Ukraine,\\ Lviv, UA--79011, Ukraine\\ }

\author{C. von Ferber}

\address{Physikalisches Institut, Theoretische Polymerphysik,
     Universit\"at Freiburg,\\  D--79104 Freiburg, Germany
}

\author{R. Folk}

\address{Institut f\"ur theoretische Physik,
Johannes Kepler Universit\"at  Linz,\\
 Linz,
A--4040, Austria }

\author{T. Yavors'kii}

\address{
Ivan Franko National University of Lviv,\\
Lviv, UA--79005,Ukraine
}
\maketitle

\pub{Received (received date)}{Revised (revised date)}

\begin{abstract}
  In this review paper, we discuss the influence of weak quenched
  disorder on the critical behavior in condensed matter and give a
  brief review of available experimental and theoretical results as
  well as results of MC simulations of these phenomena.  We
  concentrate on three cases: (i) uncorrelated random-site disorder,
  (ii) long-range-correlated random-site disorder, and (iii) random
  anisotropy.

  Today, the standard analytical description of critical behavior is
  given by renormalization group results refined by resummation of the
  perturbation theory series. The convergence properties of the series
  are unknown for most disordered models.  The main object of this
  review is to discuss the peculiarities of the application of
  resummation techniques to perturbation theory series of disordered
  models.

\end{abstract}

\section{Introduction}\label{I}
There are two main questions we want to discuss in this paper.  The
first question is a physical one: how does disorder influence
criticality?  Taken that a non-disordered (``ideal'') system possesses
a phase transition and a critical behavior governed by some scaling
laws, will these universal laws be altered by disorder introduced into
the system? The second question elaborated below is related to the unknown
convergence properties of the series that result from the perturbative
treatment of the corresponding models.

The first question we address here is rather specific since it concerns
properties in a narrow region of the phase diagram in the vicinity of a
critical point. Nonetheless, in order to give an  extensive description of
current studies in the field we are forced to restrict the subject further.
In particular, in the present paper we will be interested only in static
critical properties leaving the bulk of dynamic phenomena outside of our scope.
We will choose  three-dimensional systems of infinite size ignoring
the influence of surfaces on the critical behaviour though such problems are
very attractive \cite{Forgacs91}.  Furthermore, we
will address the subtle limit of weak disorder where the random
system can be treated as perturbed with respect to an ``ideal''
one; this disregards the class of phenomena relevant
in the strong disorder limit, like those appearing in the vicinity  of
a percolation threshold. Finally, we will touch only the quenched
``frozen'' disorder.  To summarize, the main scope of our review will
cover the static criticality of random quenched three-dimensional
systems. In spite of our restriction of the subject, a lot
of fascinating and still unsolved problems remain as we will try to
convince the reader in our subsequent exposition.

To get a quantitative description of the phenomena occurring in
the vicinity of the critical point it is standard to rely on
renormalization group methods \cite{rgbooks}. In this approach the
physical quantities describing the critical behavior are
obtained in the form of perturbation theory series.  For some
ideal (undiluted) systems these series have proven to be
asymptotic and to have a zero radius of convergence. Therefore,
special resummation methods are used to extract reliable data from
the series. However, so far there is no proof of
the asymptotic nature of the perturbation theory
series for disordered (diluted) models.
Moreover, there exist serious indications of possible
non-asymptotic divergences. Nonetheless, the bulk of results that
were obtained from the perturbation theory expansions of
disordered models are presented {\em as if} they were asymptotic.
Here arises the second question we address in this paper: how
does the possible non-asymptotic nature show up in the analysis of
perturbation theory expansions for the models with weak quenched
disorder? Putting the first and the second questions together, the
main goal of the current review will be to discuss what
information about the influence of weak quenched disorder on the
critical behavior can one extract working with renormalization
group perturbation theory series.

The set-up of the paper is as follows. In the next Section
\ref{II} we discuss different models of weak quenched disorder and
introduce the physical quantities used for their description.
Here, of main interest for us will be systems with non-correlated
point-like defects, systems with long-range correlated disorder
and systems with disorder in the form of random anisotropy. In
Section \ref{III} we give examples of experimental realizations of
such systems. These include: site--diluted magnets,
${\rm He^4}$ in porous medium, magnets with
extended impurities, polymers in random media, some amorphous
rare-earth--transition metal alloys. The effective Hamiltonians
for the disordered models under consideration are given in Section
\ref{IV} together with the main relations of the field theoretical
renormalization group theory. In Section \ref{V} we describe the
methods of resummation of divergent series and in Section \ref{VI}
we display the results obtained by their application to the field
theoretical renormalization group expansions. We conclude and give
an outlook in the Section \ref{VII}.

\section{Different types of weak quenched disorder}\label{II}

\begin{figure}[htbp]
\begin{center}
\includegraphics[width=50mm,bb= 46 515 323 784,clip]{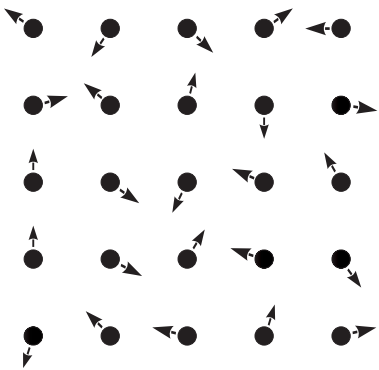}
\end{center}
\caption{The Stanley model describes a system of $m$-component
vector ``spins"   occupying sites of a $d$--dimensional
(hyper)cubic lattice and interacting by means of a short-range
translationally-invariant  force.} \label{Stanley.eps}
\end{figure}

We start our consideration from one of the simplest and most
carefully studied models which is introduced as a generalization
of the hierarchy of the lattice Ising model, the $XY$--model and the
Heisenberg model. Consider a simple (hyper)cubic lattice of the
dimension $d$, to each site of which it is prescribed a
$m$--component vector $\vec{S}$ with a fixed length (for
convenience of further notation one usually sets
$|\vec{S}|=\sqrt{m}$). Imposing a pair interaction with the energy
proportional to the scalar products between pairs of spins, one
defines the Stanley model (it is also known as $m$--vector,
$O(m)$--symmetric or generalized classical Heisenberg model). The
Hamiltonian of the Stanley model reads \cite{Stanley68}:
\begin{equation}\label{Stanley}
H =  - \frac{1}{2}\sum_{{\bf R}, {\bf R}^{\prime}} J(|{\bf R} -
{\bf R}^{\prime}|) \vec{S}_{\bf R} \cdot \vec{S}_{\bf R^{\prime}},
\end{equation}
where the $d$-dimensional vectors ${\bf R}$ span the sites of a
simple (hyper)cubic lattice (see Fig. \ref{Stanley.eps}) and
$J(|{\bf R} - {\bf R}^{\prime}|)$ is an isotropic and
translationally invariant short-range interaction. The Stanley
model has further been generalized by analytic continuation in $m$
to describe critical properties of self-avoiding walks ($m=0$),
spherical ferromagnet ($m=\infty$), the Gaussian model ($m=-2$),
etc.

A many--body system described by the model (\ref{Stanley}) can
undergo a continuous phase transition into a ferromagnetic phase.
Yet, the transition occurs only for lattice dimensions $d$ larger
than a certain lower critical dimension $d_{\rm Low}$. The
numerical value of $d_{\rm Low}$ is a function of the spin
component number $m$. If the Hamiltonian (\ref{Stanley}) remains
invariant under continuous rotations in the space of vectors
$\vec{S}$, i.e. $m\geq2$, one can show that $d_{\rm Low}=2$
\cite{d2}. Otherwise, for $m=1$ (the Ising model) the continuous
$O(m)$ symmetry degenerates into a discrete one corresponding to
the invariance of the Hamiltonian (\ref{Stanley}) under discrete
flip of all spins to the opposite direction. In this case $d_{\rm
Low}=1$ \cite{Stanley}.

In the theory of critical phenomena one is interested in the
peculiarities of the behavior in the vicinity of the critical
temperature $T_c>0$.  The critical temperature essentially depends on
microscopic details of the model. However, asymptotically close to
$T_c$ the thermodynamic and correlation functions for very different
systems may depend on the external parameters by the same power-like
functions. These functions are primarily characterized by their critical
exponents.

The critical exponent $x$ of a physical observable
${\cal O}(\tau)$ asymptotically close to the critical
point $T_c$  is defined by \cite{Stanley}:
$x \equiv \lim_{\tau\rightarrow 0} \frac{\ln {\cal
O}(\tau)}{\ln|\tau|}$, where $\tau = (T-T_c)/T_c$ is the reduced distance to
the critical point. For instance, the magnetic susceptibility $\chi$ diverges
as:
\begin{equation} \label{exp_def}
\chi \simeq \Gamma_{\pm}|\tau|^{-\gamma}\,,
\mbox{\hspace{2em}}\tau \rightarrow 0
\end{equation} where $\gamma$ is the
susceptibility critical exponent, while $\Gamma_{+}$ and $\Gamma_{-}$ are
critical amplitudes above and below the critical point respectively.  The
power law of type (\ref{exp_def}) is exact in the asymptotic regime $\tau
\rightarrow 0$.

The definitions of the most common static critical exponents  of a magnet
are provided in Table \ref{tab1}.

\begin{table}
\ttbl{32pc}{
Critical exponents governing the power-like behavior of
thermodynamic and correlation functions asymptotically close to
$T_c$. Here, $M$ is the magnetization, $\chi$ is the magnetic
susceptibility, $C_H$ is the heat capacity at constant external
magnetic field $H$; $\langle \delta M(0)\delta M({\bf R}) \rangle$
is the pair correlation function of the magnetization fluctuations and
$\xi$ is the correlation length. \protect\label{tab1}
 \newline
 }
{\begin{tabular}{l|l}
$M \propto (-\tau)^{\beta}
(T\rightarrow T_c^-,\,H=0)$
&
$ \langle\, \delta M(0) \delta M({\bf R}) \,\rangle \,\propto\,
\exp{(-|{\bf R}|/{\xi})} \,\, (T\neq T_c,\,H=0)$
\\
&\\
$\!H \propto |M|^{\delta}{\rm sgn}(M)$ ($T=T_c, H\neq 0$)
&
$\xi \propto  \tau^{-\nu}\,\,\, (T\rightarrow T_c^+,\,H=0)$  \\
&\\
$\chi \propto \tau^{-\gamma}\,(T\rightarrow T_c^+,\,H=0)$
&
$\langle\, \delta M(0) \delta M({\bf R}) \,\rangle \propto
|{\bf R}|^{-(d-2+\eta)}$ ($T=T_c, H= 0$) \\ & \\
 $C_H \propto \tau^{-\alpha}\,(T\rightarrow T_c^+,\,H=0)$
& \\
\end{tabular}}
\end{table}

In contrast to the critical temperature, the values of the critical
exponents do not depend on microscopic details of the Hamiltonian and
are determined only by the global properties of the model. For a
short-range interaction the global properties are the lattice
dimension $d$, the dimension $m$ and other symmetry properties of the
order parameter.  Due to the fact that the critical exponents may be
identical for systems with very different microscopic nature, the
critical exponents are called universal. Subsequently, if the critical
properties of two systems are described by the same set of scaling
functions and an identical set of critical exponents they are said to
belong to the same universality class.

In the case of the $O(m)$--symmetric model (\ref{Stanley})
the critical exponents are functions of the lattice and spin
dimensions $d$ and  $m$. However, if $d$ is greater than some upper critical
dimension $d_{\rm Up}=4$, the critical exponents actually do not
depend on $m$ and take mean field values $\gamma=1, \nu=1/2,
\eta=0, \beta=1/2$. For $d_{\rm Low}\leq d \leq d_{\rm Up}$
the critical exponents show non-trivial dependence on the spin
dimension $m$ and need further consideration. In particular, for
space dimension $d=3$ the critical exponents of the model
(\ref{Stanley}) have been calculated so far with very high
accuracy. The standard values of the critical exponents are listed
in the Table \ref{tab2}.

\begin{table}
\ttbl{30pc}{The standard numerical values of the critical exponents of
the $d=3$ Stanley model (results of the Ref.
\protect\cite{Guida98}).  The
cases $m=1,\,2,\,3$ correspond to Ising, $XY$- and Heisenberg models,
$m=0$ reconstitutes the so--called polymer limit.
\protect\label{tab2}
\newline
}
{\begin{tabular}{|c|c|c|c|c|c|}
\hline
$m$ & $\gamma$ & $\nu$ & $\eta$ & $\beta$ & $\alpha$ \\ \hline
0 & $1.1596\pm0.0020$ & $0.5882\pm0.0011$ & $0.0284\pm0.025$ &
$0.3024\pm0.0008$  & $0.235\pm0.003$ \\ \hline
1 & $1.2396\pm0.0013$ & $0.6304\pm0.0013$ & $0.0335\pm0.025$ &
$0.3258\pm0.0014$ & $0.109\pm0.004$ \\ \hline
2 & $1.3169\pm0.0020$ & $0.6703\pm0.0015$ & $0.0354\pm0.025$ &
$0.3470\pm0.0016$ & $-0.011\pm0.004$ \\ \hline
3 & $1.3895\pm0.0050$ & $0.7073\pm0.0035$ & $0.0355\pm0.025$ &
$0.3662\pm0.0025$ & $-0.122\pm0.010$ \\ \hline
\end{tabular}}
\end{table}

The model (\ref{Stanley}) corresponds at criticality to a wide range
of real systems, not always magnetic, and it is a convenient starting point
to study critical phenomena. However, it suffers from obvious
restrictions.  Typical for real substances are properties which can be
considered as deviations from the ``ideal'' structure of the Hamiltonian
(\ref{Stanley}). Various types of defects spoil the strict translational
invariance of the inter--spin interaction $J$ and/or can be effectively
represented by additional terms in the Hamiltonian. In both cases, disorder
introduces stochastic (random) variables in the Hamiltonian of the system.
In the following, we will treat randomness introduced by either site
dilution or random anisotropy.

\begin{figure}[t]
\begin{center}
\includegraphics[width= 50mm]{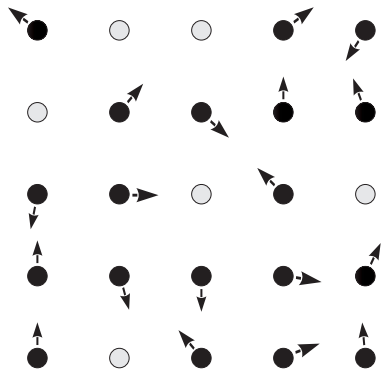}
\caption{\label{dilStan.eps} The weakly diluted Stanley model is
characterized by a small amount of randomly distributed (quenched)
non--magnetic atoms which spoil the translationally--invariant
structure of the spins in Fig. \ref{Stanley.eps}.}
\end{center}
\end{figure}

The presence of non--magnetic impurities can be modeled by a class
of site--diluted models \cite{note2}. A site--diluted Stanley
model is introduced by the Hamiltonian:
\begin{equation}\label{dilStanley}
H =  - \frac{1}{2}\sum_{{\bf R}, {\bf R}^{\prime}} J(|{\bf R} -
{\bf R}^{\prime}|) \vec{S}_{\bf R} \cdot \vec{S}_{\bf R^{\prime}}
{c}_{\bf R} {c}_{\bf R^{\prime}},
\end{equation}
where ${c}_{\bf R}$ is an occupation number: it equals $1$ when the
site is occupied by a magnetic atom and $0$ otherwise (see Fig.
\ref{dilStan.eps}). Geometrically, the vacancies $c_R=0$ can be
distributed independently according to the probability density
\begin{equation} \label{uncorr}
P(c_R)=(1-p)\delta(c_R)+p\delta(1-c_R)
\end{equation}
with $p$ being the concentration of magnetic atoms. In the absence of
correlations between the vacancies the ordered phase of the model
(\ref{dilStanley}) persists with a decrease of concentration $p$ up to
some value $p_c$. The latter depends on the lattice type and dimension
and is equal to the percolation threshold $p_{\rm perc}$ \cite{perc}.
Here, we are not interested in phenomena occurring in the vicinity of
the percolation threshold $p\sim p_{\rm perc}$ but rather we address
the limit of weak dilution $p\simeq 1$.  Since in the critical region
impurities cause, in particular, a shift of the critical temperature,
this property can be taken as a genuine property of disorder due to
non--magnetic defects. In turn, the scale of the correlation of non--magnetic
impurities becomes apparent by the fluctuations
of the local temperature of the magnetic phase transition. In what follows
below we will not consider frustrations in the exchange interaction
$J(R)$ in (\ref{dilStanley}) which may lead to the spin-glass phase as
far as the main problem of interest for us is the second order
ferromagnetic phase transition.

From the viewpoint of dynamics one may distinguish two main types of
disorder. If the dynamics of the impurities has a characteristic time
that is  comparable to the relaxation times in the pure system, the
impurity variables are treated identically to the ``pure'' dynamical
variables and are a part of the disordered system phase space. The
corresponding annealed disorder \cite{Brout59} has been subject to
numerous investigations and reveals trivial results at
criticality. The so called Fisher renormalization \cite{Fisher68}
states that when a heat capacity critical exponent of an undiluted
system $\alpha_{pure}$ is positive, then the critical exponent
$\alpha$ of the corresponding annealed diluted system changes to
$\alpha=\frac{-\alpha_{pure}}{1-\alpha_{pure}}$ and any other
critical exponent $x$ of the annealed system is determined by
the corresponding pure one ($x_{pure}$) by a simple normalization
of the form: $x = \frac{x_{_{pure}}}{1-\alpha_{_{pure}}}$. This
explains the prevailing interest in systems with quenched disorder
where the impurities can be considered as fixed. To obtain the
thermodynamics of the quenched system one needs to perform the configurational
average over an ensemble of disordered systems with different
realization of the disorder.

The influence of uncorrelated weak disorder of type (\ref{uncorr}) on
the critical behavior of the model (\ref{dilStanley}) can be predicted
by the so--called Harris criterion \cite{Harris74}. It states that
only if the heat capacity critical exponent $\alpha_{\rm pure}$ of the
pure (undiluted) system is positive, i.e.  the heat capacity diverges
at the critical point, quenched disorder causes changes of the
critical exponents. From Table I one concludes
that only the Ising model is affected by
short--range--correlated quenched disorder at criticality \cite{note}.
Moreover, the heat--capacity critical exponent of the weakly--diluted Ising
model should be negative as is suggested by the inequality $\nu \geq 2/d$
proven for diluted systems \cite{Chayes86}.

To explain possible generalizations of the model (\ref{dilStanley})
with site occupation distribution (\ref{uncorr}) let us note, that (as
it will be seen from Section \ref{V}) for well behaved distributions
the two first moments as far as they exist
determine the critical behaviour of the model. For the
distribution (\ref{uncorr}) one gets:
\begin{equation}
 \langle c_{{\bf R}}\rangle=p,
\end{equation}
\begin{equation} g(|{\bf R}-{\bf R}'|)=\langle c_{{\bf R}}c_{{\bf
R}'}\rangle - \langle c_{{\bf R}}\rangle^2=p(1-p)
\delta_{{\bf R}-{\bf R}'},
\label{point}
\end{equation} where $\langle \dots \rangle$ means averaging
with the distribution function (\ref{uncorr}), and $
\delta_{{\bf R}-{\bf R}'}$ is Kronecker's delta.

For the pair correlation function (\ref{point})  one may consider
\cite{Dorogovtsev80} more general cases.
In the first of the two generalizations treated in this review
the  quenched random
impurities are strongly correlated in $\varepsilon_d$ dimensions and
randomly distributed over the remaining $d-\varepsilon_d$ dimensions
while in a second generalization we will treat long range correlations
as described below.
In the first case, the impurities are envisaged as  parallel
$\varepsilon_d$-dimensional objects, each extending
along the coordinate directions  symbolized as ${\bf
R}_{||}$, whereas in remaining
 $d-\varepsilon_d$
dimensions symbolized as ${\bf R}_{\perp}$ they are randomly
 distributed.
The pair correlation function in this case is defined to
yield \cite{Dorogovtsev80}:
\begin{equation}
g(|{\bf R}-{\bf R'}|)=v_0\delta^{d-\varepsilon_d}
({\bf R}_{\perp}-{\bf R'}_{\perp}),
\,\,\,\,\,\mbox {with}\,v_0>0.\label{dor}\end{equation}
The case of point-like uncorrelated impurities (and the pair
correlation function (\ref{point})) is reconstituted
from (\ref{dor}) by $\varepsilon_d=0$.
Systems with impurities distribution governed by the pair
correlation function (\ref{dor}) are not covered by the
aforementioned Harris criterion. To show this,
let us note that due to fluctuations in concentration appears
the local transition temperature will vary. Denoting the deviation
from the local transition temperature by $\tau(R)$ one can
determine the mean deviation $\tau_{\xi^d}$ to the transition
temperature in a region of correlated spins,
i.e. in a volume $\xi^d$ with correlation length $\xi$, as
the average of $\tau({\bf R})$ over this region,
\begin{equation}\label{tau_xi}
\tau_{\xi^d}=\frac{1}{\xi^d}\int_{\xi^d}{\rm d}^d R \tau({\bf R}).
\end{equation}
For convenience, we pass to the continuum space $R$ with subsequent change of
Kronecker's deltas to $\delta$-functions and of sums to integrals.

The mean values (\ref{tau_xi}) vary from one correlated volume
$\xi$ to another.  One can only speak consistently of a phase
transition in the whole sample at a precise transition temperature if
these variations become negligibly small as the transition  is
approached.  The variation of the mean values $\tau_{\xi^d}$ in
different correlated regions is characterized by their variance
\begin{equation}
\Delta^2=\frac{1}{\xi^{2d}}\int_{\xi^d}{\rm d}^d R \int_{\xi^d}{\rm
d}^d R'g(|{\bf R}-{\bf R'}|).  \label{delta}
\end{equation}
For the weak impurity concentrations the shift in the local transition
temperature is proportional to the
concentration.  This allows one to substitute the pair correlation
function in (\ref{delta}) by the impurity pair correlation function
(\ref{dor}).  Evaluating  (\ref{delta}) and using  $\xi\sim
\tau^{-\nu_p}$ (with $\nu_p$ being the correlation length critical
exponent of the pure model) one finds that for $\tau\to 0$:
\begin{equation}
\Delta^2/\tau^2\sim\tau^{(d-\varepsilon_d)\nu_p-2}.
\end{equation}
This tends to zero, i.e. the variance becomes negligible, for
$   \tau\to 0 $ if the exponent of $\tau$ is positive; if it is
negative, however, the variance $\Delta$ goes to zero more slowly than
$\tau$ itself, and the mean value $\tau$ becomes meaningless
 \cite{Korzhenevskii96}. This is the indication,
that the defects qualitatively change the transition, and it leads to the
generalized Harris
criterion for the relevance of extended defects on the  critical
behaviour:
\begin{equation}\label{criterion}
\varepsilon_d>d-\frac{2}{\nu_{p}}.
\label{genhar}\end{equation}
Putting in (\ref{genhar})  $\varepsilon_d=0$ one recovers the Harris
criterion for the point-like weak quenched disorder \cite{Harris74}.

The second generalized correlation function that we treat here
describes a power-law decrease of the correlations at large distances
\cite{Weinrib}:
 \begin{equation} g(|{\bf R}-{\bf R}'|) \sim |{\bf
R}-{\bf R}'|^{-a}.  \label {gg}
\end{equation} In what follows, defects which are correlated according
to (\ref {gg}) with $a\geq d$ will be called short-range correlated,
and correspondingly, with $a<d$ - long-range-correlated. The
correlation function (\ref{gg}) has a direct physical interpretation:
straight lines of impurities of random orientation are described by
the case $a=d-1$, whereas randomly oriented planes of impurities
correspond to $a=d-2$ \cite{Weinrib}. Moreover, different non-integer
values of $a$ may be interpreted as randomly distributed impurities of
fractal nature \cite{Yamazaki88}.  Weinrib and Halperin \cite{Weinrib}
showed that in the presence of long-range power-law correlations in
the disorder the Harris criterion is also modified: for $a<d$ the
disorder is relevant, if the correlation length critical exponent of
the pure system obeys $\nu<2/a$, while for $a\geq d$ the usual Harris
criterion for the influence of short-range-correlated point defects
applies. The derivation is similar to that presented above for the
case of extended impurities.

As mentioned above, apart from site dilution as a second example
of randomness we treat a random anisotropy present at all sites.
Let us introduce a model for quenched weak random anisotropy
proposed by Harris, Plischke,
and Zuckermann \cite{Harris73}. This model describes a regular
lattice of magnetic ions, each of them being subject to a local
anisotropy field of random orientation in the form of an easy
axis. The magnitude of this field is the same for all sites. The
Hamiltonian of the random anisotropy model (RAM) then has the form
of the Hamiltonian (\ref{Stanley}) with  an additional
anisotro\-py term:
\begin{equation} \label{ram}
{\cal H} =  - \frac{1}{2}\sum_{{\bf R},{\bf R'}} J(|{{\bf R}-{\bf
R'}}|) \vec{S}_{\bf R} \cdot \vec{S}_{\bf R'}-D\sum_{{\bf
R}}({\hat z_ {\bf R}} \vec{S}_{\bf R})^2.
\end{equation}
Here, $D>0$ is an anisotropy
strength (the case of $D<0$ corresponds to the presence of easy planes in the
model), and $\hat {z}_{\bf R}$ is the unit vector pointing in the local
(quenched)  random direction of the uniaxial anisotropy.

Strong disorder at large ratio $D/J$, should destroy the second order
phase transition. For small $D/J$ this is not so obvious.  In addition
to the global variables of a regular magnet (i.e. lattice dimension,
type of interaction and spin symmetry) the low-temperature ordering in
RAM is influenced also by the distribution of the random variables
$\hat{z}\equiv\hat{z}_{\bf R}$ in (\ref{ram}).  For non-correlated
$\hat{z}$ the low-temperature ordering depends on the
probability distribution $p(\hat{z})$ of the direction of anisotropy
on a single site. As introduced by Aharony \cite{Aharony75} two kinds
of distribution functions that are relatively simple to study and that
correspond to real physical situations, are generally considered. The
first one corresponds to the isotropic case, when the random vector
$\hat z$ points with equal probability in any direction of the
$m$-dimensional hyperspace
\begin{equation}
p(\hat z)\equiv\left(\int d^m\hat z\right)^{-1}
\!=\frac{\Gamma(m/2)}{2\pi^{m/2}}, \label{dist1}
\end{equation}
with $\Gamma(x)$ being Euler's gamma-function.
The second one  corresponds to the so-called cubic anisotropy, when
the vector $\hat z$ points to any of the $2m$ directions $\hat k_i$ along the
edges of the $m$-dimensional hypercube:
\begin{equation}
p(\hat z)=\frac{1}{2m}\sum_{i=1}^m[\delta^{(m)}(\hat z-\hat k_i)+
\delta^{(m)}(\hat z+\hat k_i)]. \label{dist2}
\end{equation}
The {\em rationale} for such a choice  is to mimic the situation
when an amorphous magnet still ``remembers" the initial (cubic) lattice
structure. Other distributions may be
considered as well.

For the isotropic case (\ref{dist1}) arguments  similar to those of Imry and Ma
\cite{Imry75} applied to the random-field Ising model demonstrate
the absence of magnetization below $d=4$. Applying domain
arguments \cite{Jayaprakash80} one supposes that the ground state
will break into finite clusters of linear size $L$. This is
possible if the bulk energy $\sim DL^{-d/2}$ gained is larger than
the domain wall energy. For isotropic spin systems the latter
energy is $\sim JL^{d-2}$. This leads to the conclusion that below
four dimensions a break up into domains is favored with a
characteristic domain size $L_0\sim (J/D)^{2/(4-d)}$. Other
arguments were applied by Pelcovits {\em et al.}
\cite{Pelcovits78}. Assuming a ferromagnetic state they calculated
the correlation function of transverse magnetization and obtained
its divergence for  $d<4$. However these arguments do not consider
the anisotropic distributions $p(\hat z)$.

In the next section we introduce various physical systems
described by disordered quenched models (\ref{dilStanley}),
(\ref{ram}). A brief review of available experimental  and MC
data for these systems is given as well.

\section{Realizations of weak quenched disorder:  experiments and
Monte--Carlo simulations\label{III}}

In this section, we review critical phenomena in substances and
their computer simulation that can be considered as realizations
of the models discussed above. We start from weakly site--diluted
quenched Stanley models. Among these, the bulk of results is known
for the weakly diluted quenched Ising model as is suggested by the
Harris criterion.

\subsection{Site--diluted uniaxial antiferromagnets}\label{IIIA}

Typical experimental realizations of the weakly diluted quenched
Ising model are given  by crystalline mixtures of two compounds.
The first is an ``Ising-like'' ani\-so\-tro\-pic uniaxial
antiferromagnet with dominating short-range interaction (e.g.
${\rm FeF_2}$, ${\rm MnF_2}$), the second one is non-magnetic
(${\rm ZnF_2}$).  Mixed crystals (${\rm Fe_pZn_{1-p}F_2}$, ${\rm
Mn_pZn_{1-p}F_2}$) can be grown with high crystalline quality and
very small concentration  gradients providing an excellent
realization of random substitutional disorder of magnetic ions
(${\rm Fe^{+2}}$, ${\rm Mn^{+2}}$) by non-magnetic ones (${\rm
Zn^{+2}}$).

\begin{figure}
\epsfxsize 60mm
\centerline{\epsffile{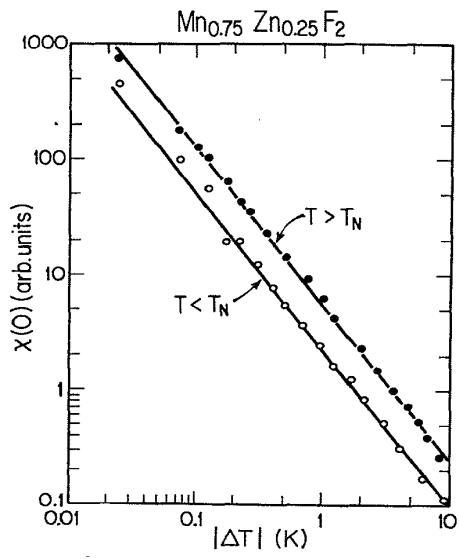}}
\vspace*{5mm}
\caption{\label{figmitchell1}
Neutron scattering measurements of the susceptibility $\chi(0)$ in
${\rm Mn_{0.75}Zn_{0.25}F_2}$ \protect\cite{Mitchell86}. The solid lines are
the results of fits to single power
laws with exponents $\gamma\simeq1.364$ above and below N\'eel temperature
$T_N$. The critical behaviour is governed by a simple power law in the
reduced temperature interval $4 \cdot 10^{-4}<|\tau|<2\cdot10^{-1}$.
}
\end{figure}

Already the first experimental studies gave evidence of new
critical behavior caused by a weak quenched dilution. In nuclear
magnetic resonance measurements of the magnetization in ${\rm
Mn_{0.864}Zn_{0.136}F_2}$ \cite{Dunlap81} the value of the
magnetization exponent $\beta= 0.349 \pm 0.008$ was found to
differ strongly from that in an undiluted sample. Within a few years
this result was corroborated by nuclear scattering measurements of
the magnetic susceptibility and correlation length critical
exponents in $\rm{Fe_pZn_{1-p}F_2}$ \cite{Birgeneau83,Belanger86}
and $\rm{Mn_pZn_{1-p}F_2}$ \cite{Mitchell86} at different
dilutions.
In particular, the latter study showed that critical
behaviour of the susceptibility in ${\rm Mn_{0.75}Zn_{0.25}F_2}$
\cite{Mitchell86} is governed by a critical exponent $\gamma$ which,
consistently with the Harris criterion, differs from that of the ``ideal''
magnet (see Fig.~\ref{figmitchell1}).
The linear birefringence measurements
revealed the cusp-like behavior of the specific heat at the transition point
with an exponent $\alpha=-0.09 \pm 0.03$ \cite{Birgeneau83}. This
experimentally proved that within the error bars the hyperscaling relation
$d\nu + \alpha = 2$ is satisfied.

Crossover phenomena in diluted systems are governed in addition to
the temperature by the concentration of the magnetic component.
The experimentally obtained exponents are often reported to be
effective ones (i.e. temperature- and dilution-dependent).
However, already in the first experiments on the critical behavior
of quenched diluted Ising-like antiferromagnets it appeared
possible to reach the asymptotic region. Thus, studying the
critical regime in Ref. \cite{Birgeneau83} the authors found
neither a region in reduced temperature $\tau$ where one finds
``pure'' Ising exponents nor any evidence of crossover from pure
to random exponents. This was explained by the crossover either
taking place outside the critical region or being too slow.  In
Ref. \cite{Barret86} the crossover from pure to diluted critical
behavior was studied and a magnetization exponent
$\beta=0.36\pm0.01$ was found that does not change under dilution
for $p \leq 0.05$. The crossover occurs within a very narrow range
of $\tau$ at relatively large values of $\tau$. In Ref.
\cite{Mitchell86} excellent agreement of the measured exponents
$\gamma=1.364\pm0.076$ and $\nu=0.715\pm0.035$ with the
theoretical values of the diluted Ising model was obtained for the
temperature range $4\cdot 10^{-4} \leq \tau \leq 10^{-1}$ and
$p=0.75$. The value of the critical exponent $\beta$ was also the
subject of a crossover analysis in Refs.
\cite{Thurston88,Rosov88}. Ref. \cite{Thurston88} concludes that
the experimental errors are too large in order to distinguish
between the pure Ising model and the diluted Ising model critical
behavior. In Ref. \cite{Rosov88} no crossover was found after the
correction-to-scaling had been taken into account, and the diluted
Ising model critical behavior was found in the whole temperature
range.

It is known that the diluted Ising magnet in a uniform magnetic field
$H$ along the uniaxial direction exhibits static critical behavior of
the random-field  Ising model \cite{ranfield}. Such experiments give
additional information about the critical behavior of the diluted Ising
model when performed for $H=0$ \cite{withH}. A comprehensive account of
the experimentally measured values of the critical exponents of diluted
uniaxial magnets is given in a recent review \cite{Folk01}.

\subsection{Disordered Heisenberg-like magnets}

Another experimental realization of the weakly diluted Stanley model
are given by magnetic systems described by the site-disordered Heisenberg
model. Some
amorphous magnets as ${\rm Fe_{90+x}Zr_{10-x}}$, ${\rm
Fe_{90-y}M_yZr_{10}}$ ( ${\rm M=Co,\,Mn,\,Ni}$)
\cite{ammagn,Perumal01}
and tran\-si\-tion-metal
based magnetic glasses \cite{magglass,Kellner86}
as well as disordered crystalline materials  as
${\rm Fe_{100-x}Pt_x}$ \cite{Boxberg94}, ${\rm Fe_{70}Ni_{30}}$
\cite{Kellner86} and ${\rm
Eu}$-chalcogenide solid solution systems (${\rm
Eu_{x}Sr_{1-x}S_{0.5}Se_{0.5}}$, ${\rm Eu_{x}Sr_{1-x}S}$, ${\rm
Eu_{x}Ba_{1-x}S}$, ${\rm Eu_{x}La_{1-x}S}$)
\cite{Westerholt83,Westerholt84,Westerholt85,Westerholt87}
belong to this group of
systems.

\begin{figure}[htbp]
\begin{center}
\includegraphics[width=70mm,bb= 54 516 492 757,clip]{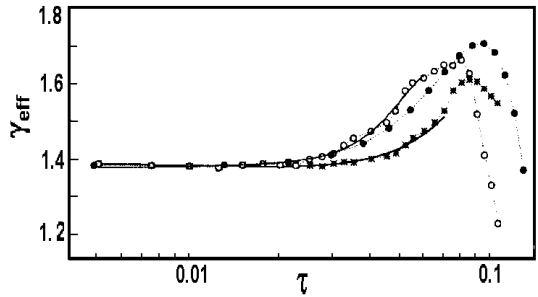}
\end{center}
\caption{Experimentally measured isothermal susceptibility effective
critical exponent $\gamma_{eff}$ for ${\rm
Fe_{90-x}Mn_xZr_{10}}$ for different values of $x$ \protect\cite{Perumal01}
($\tau=(T-T_c)/T_c$).}
\label{effect}
\end{figure}

According to the Harris criterion the critical behaviour of the diluted
Heisenberg model should remain unaltered in its asymptotics with critical
exponents of the $d=3$ pure Heisenberg model (see the last line of Table
\ref{tab2}).  However, the experiments give a wide scattering of the
values of the critical exponents. Often the reason is that the asymptotic
region still is not reached and only the effective exponent is observed. In
particular, the value of the magnetic susceptibility effective
critical exponent $\gamma_{eff}$ with decrease of distance to the critical
point goes through a peak to its asymptotic value which coincides with
magnetic susceptibility critical exponent of pure Heisenberg
magnets. The dependence of $\gamma_{eff}$ on reduced temperature
$\tau$ is presented in Fig.\ref{effect} taken from the recent paper
\cite{Perumal01}. The effective behaviour also was studied by Monte
Carlo simulations \cite{Fahnle}, the results show that
$\gamma_{eff}$ has a maximum while approaching the asymptotic regime.
Such form of temperature dependence of effective susceptibility
critical exponent is corroborated by recent theoretical investigations
\cite{Lviv}.

\subsection{${\rm He^4}$ in a porous medium}\label{IIIB}

The listed experimental evidences of the critical behavior of the
weakly diluted quenched Ising model confirm the Harris criterion.
However, the bulk of the data was obtained for the magnetic phase
transition. Due to the universal character of the critical
exponents the model (\ref{Stanley}) can describe universal
properties of critical phenomena of completely different
microscopic nature. An example is the transition of liquid
helium-4 to the superfluid state which belongs to the Stanley
model universality class at $m=2$. The uncorrelated point--like
disorder (\ref{uncorr}) in the case of a fluid can be introduced
by a porous medium \cite{note4}.
 Recently, high precision experiments on the
critical behavior of liquid helium-4 near the superfluid
transition in a porous medium \cite{He4dil} confirmed the
irrelevance of quenched disorder, again in correspondence with the
Harris criterion, since the specific heat exponent near the
superfluid transition is slightly negative \cite{note3}. More data
about effect of disorder on the $\lambda$-transition of ${\rm
He^4}$ may be found in recent review \cite{Pelissetto00}.

\subsection{MC studies of random Ising model}\label{IIIC}

Another way to study the critical behavior of the model
(\ref{dilStanley}) is to use computer Monte-Carlo experiments.
However, in contrast to experiments, the first studies of critical
behaviour of the weakly diluted Ising model on a simple cubic
lattice \cite{Landau80} revealed critical exponents that were
identical within the numerical error with the corresponding
exponents of the pure (undiluted) system within a wide dilution
region. Later, these data were objected by subsequent MC
simulations on larger lattices \cite{Marro86}, that obtained
critical exponents varying continuously with the magnetic sites
concentration $p$.  Indications of a change of the order parameter
critical exponent $\beta$ upon dilution initiated an extension of
the studies to determine the other critical exponents and to check
the scaling for disordered systems. Finally, the Swedsen-Wang
algorithm that avoids the critical slowing down of the relaxation
was applied to the $d=3$ diluted Ising model \cite{Wang89} and
resulted in critical exponents for the susceptibility and
correlation length that were independent of concentration over a
wide range of dilution.

Due to Refs. \cite{Wang89,Wang90} and especially
\cite{Heuer90,Heuer93} it became clear that the concentration
dependent critical exponents found in previous MC simulations are
effective ones, characterizing the approach to the asymptotic
region. The effective exponents $\gamma,\beta$ and $\zeta=1-\beta$
(the last one describes the divergence of the magnetization-energy
correlation function) were shown \cite{Heuer90} to be
concentration dependent in the concentration region $0.5 \leq p <
1$.  These data were refined three years later \cite{Heuer93}
resulting in more accurate estimates for the above mentioned
exponents and the critical exponent $\nu$ of the correlation
length with continuously varying values.

\begin{figure}[]
\epsfxsize 60mm
\centerline{\epsffile{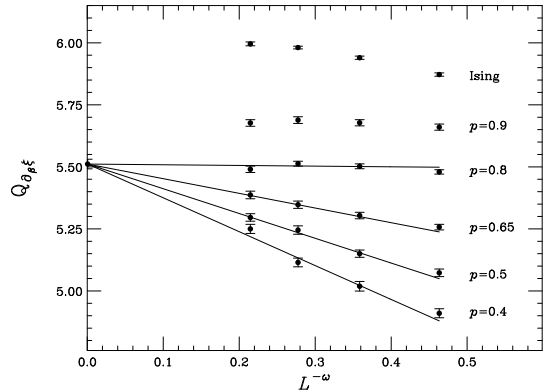}}
\vspace*{5mm}
\caption{\label{figparisi} Determination of the
correction-to-scaling exponent $\omega$ in the MC simulations  of
Ref. \protect\cite{Parisi98}. Quantity
$Q_{\partial_{\beta}\xi}=2^{1+1/\nu}$ is plotted for different
dilutions $0.4\leq p\leq 1$ and lattice sizes $8\leq L\leq 128$.
One can see that at $p=0.9$ the system is crossing over from the
pure Ising fixed point to the diluted one even for $L=128$. The
solid lines correspond to a fit $\omega=0.37$ yielding the same
infinite volume extrapolation for all $p\leq 0.8$. }
\end{figure}

The critical behavior of the $d=3$ diluted Ising model was
reexamined recently in Ref. \cite{Parisi98}. The study was based
on the crucial observation that it is important to take into
account the leading correction-to-scaling term in the infinite
volume extrapolation of the MC data.
The simulations confirmed
the universality of the critical exponents of the $d=3$ diluted
Ising model over a wide region of concentrations
(see Fig.~\ref{figparisi})
. In particular the value of the
correction-to-scaling exponent $\omega$ was found to be
$\omega=0.37\pm 0.06$ which is almost half as large as the
corresponding value in the pure $d=3$ Ising model $\omega=0.799\pm
0.011$ \cite{Guida98}. The smallness of $\omega$ in the dilute case
explains its importance for an analysis of the asymptotic critical
behavior. The other exponents were determined as:  $\beta=0.3546 \pm
0.0028$, $\gamma=1.342 \pm 0.010$, $\nu=0.6837 \pm 0.0053$. We refer
the interested reader to the papers \cite{Folk00,Folk01} where the
numerical values of the exponents obtained in the MC simulations are
listed.

\subsection{Magnets with long-range-correlated quenched disorder}

\label{IIID} Whereas at present many experimental and MC
simulation data are available for magnetic systems with
random-site uncorrelated disorder, the critical behavior of
systems with long-range-correlated disorder is much less
investigated. Therefore, in this subsection that is devoted to
systems with this type of disorder we present only a few works
where its influence has been studied.

In magnets, long-range-correlated disorder may be present in the form
of continuously distributed dislocations and disclinations,
so-called extended structural defects. These defects may have the form
of lines or planes of random orientation \cite{Yamazaki} or may form
some sponge-like fractal objects, which are considered as aggregation
clusters \cite{Yamazaki88}. The particular case of systems with
dislocation lines or planes of parallel orientation, is investigated in Ref.
\cite{Cardy}.

Recently, Ballesteros and Parisi \cite{Ballesteros99} presented
Monte-Carlo simulations of the diluted Ising model in three
dimensions with extended defects in the form of lines of parallel
orientation. Finite-size scaling techniques were used to compute
the critical exponents of these systems, taking into account the
strong scaling corrections. The obtained value for the exponent
$\nu$ is compatible with the analytic predictions \cite{Weinrib}.
In Ref. \cite{Marques00} the three-dimensional so-called thermally
diluted Ising systems were studied in Monte-Carlo simulations.
Here, at the characteristic ordering temperature $\theta$, the
spins of the dominant type (concentration $p\geq 0.5$) are taken
as the location of the magnetic atoms, and the rest are taken as
magnetic vacancies. The structure of the realization is fixed
thereafter for all temperatures at which the magnetic interactions
are subsequently investigated. If $\theta$ happens to coincide
with the critical temperature of the undiluted Ising model, the
vacancies are in randomly located points, but with a
long-range-correlated distribution.  It was found that the
simulated critical exponents of thermally diluted Ising systems at
criticality agree fairly well with theoretical predictions of
Weinrib and Halperin \cite{Weinrib}.

We adduce here also the interesting result of Monte-Carlo
study of three-di\-men\-si\-onal Ising model
with bond disorder, which is random
in the vertical direction and  correlated in $\varepsilon_d=2$
horizontal directions \cite{Lee92}.
The results demonstrate a sharp second-order phase transition, with
exponents of new universality class. However, it seems not to be of the same
universality class as the model of extended defects presented by
Dorogovtsev \cite{Dorogovtsev80}.

\subsection{Polymers in random media}\label{IIIE}

Another example of critical behavior  in the presence of weak
quenched disorder is found for the scaling properties of long
flexible polymer chains in random media. The question of how linear
polymers behave in disordered media is not only interesting from a
theoretical point of view, but is also relevant for understanding
transport properties of polymer chains in porous media, such as
enhanced oil recovery, gel electrophoresis, gel permeation
chromatography etc.

It is established that the universal scaling properties of long
flexible polymer chains in a good solvent are perfectly described
within a model of self-avoiding walks (SAWs) on a regular lattice
\cite{desCloizeaux90}.  Their scaling properties in the limit of
an infinite number of steps are governed by the same scaling laws
as the pure (undiluted) Stanley model at the critical point in the
formal $m\rightarrow 0$ limit \cite{deGennes79}. In particular,
for the average square end-to-end distance $\langle R^2 \rangle$
and the number of configurations $Z_N$ of a SAW with $N$ steps on
a regular lattice one finds in the asymptotic limit $N\to\infty$:
\begin{equation}\label{nu} \langle R^2 \rangle \sim
  N^{2\nu},\mbox{\hspace{3em}} Z_N \sim {\tilde z}^{N} N^{\gamma-1}
\end{equation} where $\nu$ and $\gamma$ are the universal correlation
length and susceptibility exponents for the $m=0$ model that only
depend on the space dimensionality $d$ and $\tilde{z}$ is a
non-universal fugacity.  For $d=3$ the exponents are given in Table 1.

The problem of SAWs on randomly diluted lattices, which may serve as a
model of polymer chains in a porous medium, has been the subject of
intensive discussion
\cite{Chakrabarti,Harris83,Kim,Meir89,Grassberger93,Barat95,randomsaws,Ordemann}.
The
main question of interest here is: does a small amount of quenched
structural point-like defects in the medium induce changes to the
universal properties of the polymer macromolecules? One should be
careful in applying the Harris criterion ``naively" to the SAW
problem.  Although the critical exponent $\alpha$ of a SAW on the
$d=3$ dimensional pure lattice is positive \cite{Guida98}
($\alpha(d=3)=0.235\pm0.003$), a weak quenched short-range-correlated
disorder does not alter the SAW critical exponents. This statement has
been proven by Harris \cite{Harris83} and confirmed later by
renormalization group results \cite{Kim}.  Numerical results for these
systems available from Monte-Carlo simulations also lead to the conclusion
that such disorder does not affects the scaling behavior of SAWs
\cite{Barat95,ransaws}.

On the other hand, universal properties of SAWs on a percolation
cluster obey new scaling laws (see, e.g. \cite{Meir89,Blavats'ka01a}
and references therein).

Another interesting problem is the scaling behavior of polymers in
disordered media when the defects are correlated or belong to some
sponge-like structure. In Refs. \cite{Blavats'ka01a,Blavats'ka01}
it was found that the asymptotic behavior of SAWs in long-range-correlated
disorder of type (\ref{gg}) is governed by a set of critical exponents
which are different from that in the pure case.  This recent result
waits for its experimental verification.

\subsection{Random anisotropy magnets}\label{IIIF}

Among disordered materials containing magnetic elements there are
so\-me intermetallic amorphous alloys, which are characterized not
by a random-site disorder, but by a disorder in the form of a
random anisotropy. Random anisotropy is the most important
characteristic of these metals and that results from their random
close-packing structure. Typical examples of random anisotropic
materials are amorphous rare-earth--transition metal alloys
\cite{Harris73}. Their features are taken into account by the
random anisotropy model (RAM) in (\ref{ram}). The model may be
extended also to describe other intermetallic amorphous alloys
containing rare-earth elements with asymmetric charge distribution
\cite{Cochrane78}.

The original example for the RAM was ${\rm TbFe_2}$
\cite{Harris73}. Some of its structural features were used for
testing the randomly packed hard spheres model corresponding to
rare-earth--transition metal alloys. The ${\rm Tb-Tb}$ correlation
function, obtained on the basis of generated clusters with spheres
of two sizes in Ref. \cite{Cochrane74} is in qualitative agreement
with neutron scattering data on amorphous ${\rm TbFe_2}$. Using
the simple charge model, both classical \cite{Cochrane74} and
quantum mechanical calculations \cite{Cochrane75} carried out on
these clusters show the evidence of the anisotropy term in
(\ref{ram}) as well as  that the distribution of directions of the
local easy axes is effectively isotropic, and indicate the
apparent absence of correlation between the directions of axes at
neighboring sites.

Magnetic properties of the surroundings of rare-earth atoms are neglected
within this model and thus the RAM does not describe correctly alloys of
rare-earth -- ${\rm Fe}$ type. Realistic models would have to include
both magnetic iron-iron  and rare-earth -- iron interactions.
Therefore the RAM is most adequate to describe the intermetallic alloys
of rare-earth with non-magnetic compounds.

A considerable number of various experimental data on alloys described by
the RAM are listed in an exhaustive review by Cochrane {\em et al.} \cite{Cochrane78}.
As follows from their analysis, some of these materials order magnetically
at sufficiently low temperature, but the nature of this ordering is not
clear. Among possible low--temperature phases they discuss
ferromagnetic ordering, spin--glass phase, quasi long--range ordering.
It is clear, that the type of ordering is connected with the alloy intrinsic
structure, that corresponds within the model to the distribution of anisotropy
axes.

Another interesting application of the RAM is the description of the
isotropic-nematic phase transition in liquid crystals in porous media.
Monte Carlo simulations give the quasi long-range order in this case
\cite{Chakrabati98}, while experimental results suggest that the bulk
nematic phase is replaced by a ``glassy" state \cite{Wu92}.

\section{Effective Hamiltonians and the renormalization group}\label{IV}

In the theory of critical phenomena it is standard to rely on the
methods of renormalization group (RG). The basic conjecture of this
approach states that a system undergoing a second--order phase
transition is scale--invariant at the critical point $T_c$.  Near to
the critical point the behavior of thermodynamic and correlation
functions are determined only by the correlation length $\xi$ (c.f.  Table
I) which describes the scale of the correlations of the order parameter
fluctuations.

Among the various mathematical schemes of renormalization group the
most accurate numerical results for critical phenomena
are obtained in terms of the field--theoretical formulation
\cite{rgbooks}. In this formulation the universal critical properties
of a statistical system are obtained from the large--scale behavior
of the corresponding effective field--theoretical Hamiltonian.  The
latter is treated with the well--elaborated methods of quantum field
theory. The relevant global properties of a microscopic model is
represented by the structure of the effective Hamiltonian.  In
particular, the effective Hamiltonian that describes the critical
behavior of the Stanley model (\ref{Stanley}) reads
\begin{equation} \label{phi4}
{\cal H}(\vec{\phi})=\int {\rm d}^d x \Big\{ {1\over 2}
\left[|\nabla \vec\phi|^2+ \mu_0^2 |\vec\phi|^2\right] + {\tilde
u_0\over 4!} |\vec\phi|^4\Big\}.
\end{equation}
and is usually referred to as the Hamiltonian of the
Landau-Ginzburg-Wilson theory. Here, $\mu_0^2$ is a bare mass
proportional to the temperature difference $\tau$ to the critical
point, $\vec\phi=\vec\phi({\bf R})$ is a $m$--component vector field,
$\tilde u_0$ is a bare coupling. Note, that the effective Hamiltonian
(\ref{phi4}) preserves the $O(m)$ symmetry of the Stanley model
(\ref{Stanley}).

\subsection{Weakly diluted Stanley model with point--like
uncorrelated defects}\label{IVA}

A weak quenched disorder term can be introduced directly into the
effective Hamiltonian (\ref{phi4}). As it was mentioned in Section
\ref{II} the presence of non--magnetic impurities in a microscopic
model (\ref{dilStanley}) manifests itself in fluctuations of the local
temperature of the phase transition.  Introducing $\psi=\psi({\bf R})$
as the field of local critical temperature fluctuations, one obtains
the effective disordered Hamiltonian \cite{Grinstein76}:
\begin{equation} \label{phi4_psi}
{\cal H}_{\rm \psi}(\vec{\phi})=\int {\rm d}^d x \Big\{ {1\over 2}
\left[|\nabla \vec\phi|^2+ (\mu_0^2+\psi) |\vec\phi|^2\right] +
{\tilde u_0\over 4!} |\vec\phi|^4\Big\}.
\end{equation}
The Hamiltonian (\ref{phi4_psi}) depends on a number of macroscopic
parameters that describe the specific configuration of the field
$\psi({\bf R})$.  On the other hand, the observables should not depend
on the specific realization of the random field $\psi$ and are to be
averaged over the possible configurations of $\psi$ \cite{Brout59}.  In
particular, the singular contribution to the free energy of the
diluted quenched Stanley model (\ref{dilStanley}) can be written in
the form of a functional integral:
 \begin{equation}
\label{conf_average}
F \propto \int D[\psi(x)] P[\psi(x)] \ln Z
[\psi(x)],
\end{equation}
where the configurationally--dependent partition function $Z[\psi(x)]$
is the normalizing factor of the Gibbs distribution with effective
Hamiltonian (\ref{phi4_psi}); $P[\psi]$ defines a probability
distribution of the field $\psi(x)$.

In order to avoid the averaging  of the logarithm in (\ref{conf_average}) the
standard step is to use the replica trick \cite{replicas}. This amounts to
writing the logarithm in the form of the following limit:
\begin{equation}
\label{trick}
\ln Z = \lim_{n\rightarrow 0}\frac{Z^n-1}{n}.
\end{equation}
While the powers of $Z$ can only be evaluated for integer values of $n$,
analytical continuation in $n$ is assumed to perform the limit $n\to 0$.
Introducing $n$ replicas of the model (\ref{phi4_psi})
and taking that $\psi$ obeys a Gaussian distribution:
\begin{equation}
\label{psi_distribution}
P(\psi)=\frac{1}{\sqrt{4\pi}w}\exp(-\psi^2/4\omega^2)
\end{equation}
with $\omega^2$ being the dispersion parameter, one ends up
\cite{Grinstein76} with the effective Hamiltonian:
\begin{eqnarray}\label{dil_phi4}
{\cal H}(\vec{\phi}){=}\! \int\! {\rm d}^d x \Big\{
{1\over 2} \sum_{\alpha=1}^{n} \left[|\nabla \vec{\phi}^\alpha|^2{+}
\mu_0^2 |\vec{\phi}^\alpha|^2\right]{+}
{u_{0}\over 4!}
\sum_{\alpha=1}^{n}\left(|\vec{\phi}^\alpha|^2 \right)^2 {+}
{v_{0}\over 4!}
\left(\sum_{\alpha=1}^{n}|\vec{\phi}^\alpha|^2 \right)^2\!
\Big\}.\nonumber\\
\end{eqnarray}
In the limit $n\rightarrow 0$ the field theory (\ref{dil_phi4})
describes the critical properties of the weakly diluted Stanley
model. Here, the bare coupling $u_0$ is positive (being
proportional to $\tilde u_0$) whereas the bare coupling $v_0$ is
proportional to minus the variance of the random variable $\psi$
and thus it is negative. The last term in (\ref{dil_phi4}) is
present only for non-zero dilution: it is directly responsible for
the effective interaction between replicas due to the presence of
impurities \cite{Dimo}.

The model (\ref{dil_phi4}) is interesting also in the polymer
limit $m\rightarrow 0$. In this case it can be interpreted as a
model for SAWs in disordered media.  Note, that this limit is not
trivial. As noticed by Kim \cite{Kim}, once the limit $m,n\to 0$
has been taken, both $u_0$ and $v_0$ terms are of the same
symmetry, and an effective Hamiltonian with one coupling
$U_0\equiv u_0+v_0$ of $O(mn=0)$ symmetry results.  This leads to
the conclusion that  weak quenched uncorrelated disorder does not
change the universal critical properties of SAWs.

\subsection{Weakly diluted Stanley model with long--range correlated
defects} \label{IVB}

In the above subsection we obtained an effective Hamiltonian for
systems with uncorrelated point-like disorder.  Now let us consider
the case, when the impurities are long-range-correlated
\cite{Weinrib}.  In particular, let us consider the case, when the
inhomogeneities in the system cause fluctuations in the local
transition temperature $T_{\rm c}(\vec{x})$, characterized by the
correlation function $g(|\vec{x}-\vec{y}|)= \langle T_{\rm
  c}(\vec{x})T_{\rm c}(\vec{y}) \rangle - \langle T_{\rm c}(\vec{x})
\rangle^2$, that falls off with distance according to the power law
(\ref{gg}). The Fourier-transform $\tilde g(k)$ of $g(x)$ reads for
small $k$:
\begin{equation}
\tilde g(k)\sim v_0+w_0k^{a-d}.
\label{furie}
\end{equation}
Note, that in the case of random uncorrelated point-like defects the
site-occupa\-tion correlation function reads: $ g(|\vec{x}-\vec{y}|)\sim
\delta(\vec{x}-\vec{y}),$ so its Fourier transform obeys:
\begin{equation}
\tilde g(k)\sim v_0.
\label{uncor}
\end{equation}

As above, we apply the replica trick to average the free energy over
different configurations of quenched disorder and construct the
effective Hamiltonian of the Stanley model with
long-range-correlated disorder \cite{Weinrib}:
\begin{eqnarray} \nonumber
{\cal H}(\vec{\phi})&=&\int {\rm d}^d x \Big\{ {1\over 2}
\sum_{\alpha=1}^{n} \left[|\nabla \vec{\phi}^\alpha|^2+ \mu_0^2
|\vec{\phi}^\alpha|^2\right]+ {u_{0}\over 4!}
\sum_{\alpha=1}^{n}\left(|\vec{\phi}^\alpha|^2 \right)^2\Big\}+ \\
&&{}\sum_{\alpha,\beta=1}^{n} \int{\rm d}^dx{\rm d}^dy
g(|\vec{x}-\vec{y}|)
|\vec{\phi}^\alpha(x)|^2|\vec{\phi}^\beta(y)|^2. \label{hamlr}
\end{eqnarray}
Here, the notations are as in formula (\ref{phi4}),
and the replica interaction vertex $g(x)$ is the
correlation function with Fourier image (\ref{furie}).

Passing  to the Fourier image in (\ref{hamlr}) and taking into
account eq.(\ref{furie}), an effective Hamiltonian results that contains
three bare couplings $u_0,v_0,w_0$.  Note, that as far as we are
interested in the long-wave limit $k\to 0 $, for $a>d$ the $w_0$-term
in (\ref{furie}) becomes irrelevant and one obtains an effective
Hamiltonian of a quenched diluted (short-range correlated) Stanley
model \cite{Grinstein76} with two couplings $u_0,v_0$.  For $a<d$ we
have, in addition to the momentum-independent couplings, the momentum
dependent one $w_0k^{a-d} $.
Note, that magnetic systems
with parallel extended impurities are not mentioned in our following
description; we refer the interested reader to the Ref.  \cite{acta}
for review and numerical results.

The analysis of SAWs in media with long-range-correlated disorder
uses Kim's observation \cite{Kim} that in the limit $m,n \rightarrow 0$ one
may pass to an effective Hamiltonian with only two couplings $U_0=
u_0+v_0$ and $w_0$ (in what follows below we will keep the notation
$u_0$ for this new coupling $U_0$).  In discrete momentum space this
effective Hamiltonian reads  \cite{Blavats'ka01a,Blavats'ka01}:
\begin{eqnarray} \label{saw}
&&{\cal H}(\vec{\phi})=\sum_{k}\sum_{\alpha}^n \frac{1}{2}
(\mu_0^2+k^2)(\vec{\phi}_k^{\alpha})^2 +
\frac{u_0}{4!}\sum_{\alpha}^n\sum_{k_1 k_2 k_3k_4}\!\!\!\!
\delta(\sum_{i=1}^4k_i)
(\vec{\phi}_{k_1}^{\alpha}\vec{\phi}_{k_2}^{\alpha})
(\vec{\phi}_{k_3}^{\alpha}\vec{\phi}_{k_4}^{\alpha}) + \nonumber\\
&& \frac{w_0}{4!}\sum_{\alpha \beta}^n
\sum_{kk_1k_2k_3k_4}\!\!\!\!|k|^{a-d}\delta(\sum_{i=1}^2k_i+k)
\delta(\sum_{i=3}^4k_i-k)\! (\vec{\phi}_{k_1}^{\alpha}
\vec{\phi}_{k_2}^{\alpha})\!
(\vec{\phi}_{k_3}^{\beta}\vec{\phi}_{k_4}^{\beta}),
m,n \rightarrow 0.\!
\end{eqnarray}
Here, the $\delta(k)$ represent products of Kronecker symbols and
the notation $( \vec{\phi}\vec{\phi})$ implies a scalar product.
Note that the $w_0$-term contains the interactions between
replicas and an additional power of an internal momentum. Again,
it may be shown that for $a=d$  in the limit $m,n \rightarrow 0$
both $u_0$ and $w_0$ terms are of the same symmetry and one is
left with an $O(mn=0)$-vector model with only one coupling
$(u_0+w_0)$ \cite{Blavats'ka01}.

\subsection{Stanley model with random anisotropy} \label{IVC}

In the cases considered so far, point-like defects introduced
random scalar variables to the effective Hamiltonian (\ref{phi4}).
Random anisotropy disorder on the other hand (\ref{phi4}) is
different in this respect. The random variable of the disordered
effective Hamiltonian for the RAM has the form of a vector with
random orientation  and given magnitude. Assuming this magnitude
as constant one  obtains the replicated Hamiltonian as follows
\begin{eqnarray}
\label{effram}\!\!\!\!\!{\cal H}_{\rm\hat z}(\vec{\phi}) &{=}\!\!& \int\! {\rm
d}^dx \left\{{1\over 2} \sum_{\alpha=1}^{n}\left[|\nabla
\vec{\phi}^\alpha|^2{+}\mu_0^2 |\vec{\phi}^\alpha|^2\right]\!{+}
\frac{u_0}{4!} \sum_{\alpha=1}^n|\vec{\phi}^\alpha|^4{-}\!
\sum_{\alpha=1}^n D\!\left( \hat z
\vec{\phi}^\alpha\right)^2\right\}.
\end{eqnarray}
Here, the  notations are as in (\ref{phi4}), $D$ is the magnitude of the
anisotropy field, $\hat z$ is a random unit vector pointing in the
direction of the local anisotropy field.

As it was noted above, to perform the average over the random
variables $\hat z$ we will use the distributions (\ref{dist1}) and
(\ref{dist2}) given in Section \ref{II}. For the isotropic
distribution of the directions of random aniso\-tro\-py averaging
the axis of all sites (\ref{dist1}) results in an effective
Hamiltonian of the following form \cite{Aharony75}
\begin{eqnarray}
\label{hamis}
{\cal H}(\vec{\phi})&{=}& \int {\rm d}^d x \Biggl\{
{1\over 2} \sum_{\alpha=1}^{n} \left[|\nabla \vec{\phi}^\alpha|^2+
\mu_0^2 |\vec{\phi}^\alpha|^2\right]+ {u_{0}\over 4!}
\sum_{\alpha=1}^{n}|\vec{\phi}^\alpha|^4 +\Biggr. \nonumber
\\ &&
\left.{v_{0}\over 4!}
\left(\sum_{\alpha=1}^{n}|\vec{\phi}^\alpha|^2 \right)^2
+
\frac{w_0}{4!}\!\sum_{\alpha,\beta=1}^n\sum_{i,j=1}^{m}
\!\phi_i^\alpha\phi_j^\alpha\phi_i^\beta\phi_j^\beta
\right\},
\end{eqnarray}
with the bare couplings $u_0>0$, $v_0>0$, $w_0<0$. Furthermore,
the values of $v_0$ and $w_0$ are related to $D$ and appropriate
cumulants of the  distribution function (\ref{dist1}) in such a
way that their ratio equals $w_0/v_0=-m$. Note that the symmetry
of the $u_0$ and $v_0$ terms corresponds to the random site
Stanley model (\ref{dil_phi4}). However the $v_0$-term has the
opposite sign.

Now, we consider the case of the distribution  of $\hat z$ along
cubic axes (\ref{dist2}). After performing the averaging over
configurations the corresponding effective Hamiltonian reads
\cite{Aharony75}:
\begin{eqnarray}
{\cal H}(\vec{\phi})&{=}& \int {\rm d}^d x \left\{ {1\over 2}
\sum_{\alpha=1}^{n}\right.\left[|\nabla \vec{\phi}^\alpha|^2+
\mu_0^2 |\vec{\phi}^\alpha|^2\right]+ {u_{0}\over 4!}
\sum_{\alpha=1}^{n}|\vec{\phi}^\alpha|^4 + \nonumber
\\
&&
\left.{v_{0}\over 4!}
\left(\sum_{\alpha=1}^{n}|\vec{\phi}^\alpha|^2 \right)^2+
\frac{w_0}{4!}\sum_{i=1}^m\sum_{\alpha,\beta=1}^n
{\phi_i^\alpha}^2{\phi_i^\beta}^2{+}
\frac{y_0}{4!}\sum_{i=1}^m\sum_{\alpha=1}^n{\phi_i^\alpha}^4\right\},
\label{hamcub}
\end{eqnarray}
where the bare couplings are $u_0>0$, $v_0>0$, $w_0<0$. The
symmetry of the $w_0$ terms in (\ref{hamcub}) differs from that in
(\ref{hamis}).  Furthermore, the values of $w_0$ and $v_0$ differ
for the Hamiltonians (\ref{hamis}) and (\ref{hamcub}) due to the
cubic distribution (\ref{dist2}) but their ratio equals $-m$
again. The last term in (\ref{hamcub}) combines the symmetries of
terms with coefficients $u_0$ and $w_0$. It does not result from
the functional representation of the free energy but is generated
by further application of the RG transformation. Therefore $y_0$
can be of either sign.

\subsection{Field--theoretical renormalization--group approach}
\label{IVD}

In this subsection, we give a brief account of the main relations of
the field-theoretical renormalization group (RG) formalism that can be
evaluated in different variants. We choose the massive field theory
scheme with renormalization of the one-particle irreducible vertex
functions $\Gamma_0^{(L,N)}(k_1,..,k_L;p_1,..,p_N;\mu^2_0;\{\lambda_0 \})$
at non-zero mass and zero external momenta \cite{Parisi}.
The one-particle irreducible vertex function  can be
defined as:
\begin{eqnarray} \label{vertex} && \delta(\sum k_i + \sum   p_j)
\Gamma_0^{(L,N)}(\{k\};\{p\}; \mu^2_0;\{\lambda_0 \}) =
\int^{\Lambda_0} e^{i(k_i R_i+p_j r_j)} \times
\nonumber\\  &&
\langle \phi^2(r_1)\dots\phi^2(r_L) \phi(R_1)\dots \phi(R_N)
\rangle^{{\cal  H}_{eff} }_{1PI}
{\rm d}^d R_1 \dots {\rm d}^d R_N {\rm d}^d r_1 \dots {\rm d}^d
r_L \, .
\end{eqnarray}
Here,  $\{\lambda_0\}$ stands for the set of  bare couplings
of the effective Hamiltonian,  $\{ p\}, \{k \}$ are the sets of
external momenta, $\Lambda_0$ is the cutoff, and the averaging is
performed with the corresponding effective Hamiltonian, ${\cal H}_{eff}$.

An intrinsic feature of the evaluation scheme (\ref{vertex}) are
divergences in the limit $\Lambda_0\rightarrow \infty$. In order
to remove them and to map the divergent mathematical objects to
convergent physical quantities one performs a controlled
rearrangement of the series for the vertex functions. Introducing
renormalizing factors for the fields $Z_{\phi}$, $Z_{\phi^2}$, the
renormalized vertex functions $\Gamma_R^{(L,N)}$ are expressed in
terms of the bare vertex functions as follows:
\begin{equation} \label{14a}
\Gamma_R^{(L,N)}(\{k\};\{p\};\mu^2;\{\lambda\}) =
Z_{\phi^2}^{L} Z_{\phi}^{N/2} \Gamma_0^{(L,N)}(\{k\};\{p\};\mu_0^2;
\{\lambda_0\}),
\end{equation}
where $\mu$, $\{\lambda \}$ are the renormalized mass and couplings.

The regularization scheme suggested in Eq. (\ref{14a}) is not unique.
To define the regularization scheme completely one imposes
renormalization conditions for the renormalized vertex functions. In
the massive scheme the renormalization conditions read
\cite{rgbooks,Parisi}:
\begin{eqnarray} \nonumber
\Gamma_R^{(0,2)}(k,-k;\mu^2,\{ \lambda \})|_{k=0}&=&\mu^2,\\
\nonumber
\frac{d}{dk^2}\Gamma_R^{(0,2)}(k,-k;\mu^2,\{\lambda\})|_{k=0}&=&1,\\
\nonumber
\Gamma_{R,\lambda_j}^{(0,4)}(\{k\};\mu^2,\{\lambda\})|_{k=0}&=&
\mu^{4-d}\lambda_j,\\ \label{conditions}
\Gamma_R^{(1,2)}(p;k,-k;\mu^2,\{\lambda\})|_{k=p=0}&=&1.
\end{eqnarray}
In the case of long-range-correlated disorder we have another global
parameter $a$ along with $d$, and for the renormalization of the
coupling $w$ one imposes \cite{Prudnikov}:
\begin{equation} \label{adcond}
\Gamma_{R,w}^{(0,4)} (\{k\};\mu^2,\{\lambda\})|_{k=0}=
\mu^{4-a}w
\end{equation}
with the renormalized coupling $w$. The scaling properties of the
system asymptotically close to the critical point
are expressed by the homogeneous Callan-Symanzik equations
\cite{rgbooks} for  $\Gamma_R^{(L,N)}(\{p\};\{k\};\mu^2,\{\lambda\})$:
\begin{eqnarray} \label{Callan}
\Big \{ \mu \frac{\partial}{\partial \mu} &{+}& \sum_i
\beta_{\lambda_i}(\{\lambda\}) \frac{\partial}{\partial
\lambda_i}-\nonumber\\&&
 (\frac{N}{2}-L)\gamma_\phi(\{\lambda\}){+}
L \bar{\gamma}_{\phi^2}(\{\lambda\}) \Big \}
\Gamma^{(L,N)}_R(\{p\};\{k\};\mu^2,\{\lambda\}) = 0,
\end{eqnarray}
 with the coefficients:
\begin{equation} \nonumber
\beta_{\lambda_i}(\{\lambda\})=\frac{\partial \lambda_i}{\partial
\ln \mu }|
_{\{\lambda_0\},\mu_0}, \,\,
\gamma_{\phi}=\frac{\partial Z_{\phi}}{\partial \ln \mu }|_
{\{\lambda_0\},\mu_0}, \,\,
\bar{\gamma}_{\phi^2}=-\frac{\partial \bar{Z}_{\phi^2}}{\partial \ln
\mu }|_ {\{\lambda_0\},\mu_0}
\end{equation}
where $\bar{Z}_{\phi^2}=Z_{\phi^2} Z_{\phi}$.

The fixed points $\{\lambda^*\}$ of the RG transformation are
given by the solution of the system of equation:
\begin{equation}
\beta_{\lambda_i}(\{\lambda^*\})=0,\phantom{555}i=1,2,\ldots.
\label{fp}
\end{equation}
The stable fixed point, corresponding to
the critical point of the system, is defined as the fixed point
where the stability matrix:
\begin{equation}\label{stmatrix}
B_{ij}=\frac{\partial \beta_{\lambda_i}}{\partial \lambda_j}
\end{equation}
possesses eigenvalues $\{\omega_i\}$ with positive real parts. In
this point, the function $\gamma_{\phi}$ determines the value of
the pair correlation function critical exponent $\eta$:
\begin{equation}\label{eta}
\eta=\gamma_{\phi}(\{\lambda^*\}),
\end{equation}
and the correlation length critical exponent $\nu$ is determined as:
\begin{equation}\label{nu1}
\nu^{-1}=2-\gamma_{\phi}(\{\lambda^*\})-
\bar{\gamma}_{\phi^2}(\{\lambda^*\}).
\end{equation}
All other critical exponents may be obtained from familiar scaling
laws. For example, the susceptibility exponent $\gamma$ is given
by:
\begin{equation}
\gamma=\nu(2-\eta).
\end{equation}

The $\beta$-- and $\gamma$--functions are calculated
perturbatively as series in the (several) couplings $\lambda_i$.
The order of the expansion corresponds to the number of loops in
the diagrammatic Feynman representation of the vertex functions
(\ref{vertex}). However, due to the (asymptotic) divergence of the
RG functions series, one cannot directly derive the reliable
physical information from these expressions. The methods that
permit to cope with the task will be described in the next
section.

\section{Resummation of asymptotic series}\label{V}

The expansions of RG functions in powers of couplings (weak
coupling expansions) are known for a number of models with high
accuracy. However, the very idea of perturbation theory is that
the result is successively approximated by accounting higher order
contributions. But the method seems doomed to failure if the
series are divergent. This is the case in the field--theoretical
renormalization group approach. Moreover, here the series are
characterized by a factorial growth of the coefficients implying a
zero radius of convergence. A simple method to manage this problem
is optimally truncating the series, i.e. accounting only its first
several terms that do not show the divergence. A self--consistent
way to take into account the higher order contributions requires
the application of special ways of resummation: Borel resummation
accompanied by certain additional procedures \cite{Hardy48}.
Resumming a series one should note that the accuracy of the
resummed results depends on the procedure of resummation. Thus the
analysis and appropriate selection of an adapted resummation
procedure is crucial to obtain reliable results.

However, there remains the principal question about the Borel
summability of the perturbation theory series for a given model.
Up to now this has been proved only for the $\phi^4$ theory with one
coupling \cite{borelsummability}. For years, the field theoretical
RG series for models with several couplings were analyzed {\em as
if} they are asymptotically divergent without a proof of this
property. Moreover, there exists strong evidence of possible
Borel non-summability of the series obtained for disordered
models \cite{Griffiths69,nonsum,Alvarez00}. In this section, we
will give the definition of an asymptotic series and provide the
methods which currently are used to extract convergent data.

\subsection{An asymptotic series}\label{VA}

For a function $f(g)$ given in the form of a power series
$$
f(g)=\sum_{k=0}^{\infty}a_kg^k,$$
the basic definition of its asymptotic nature is:
$$|f(g) -\sum_{k=0}^{N}a_kg^k|\leq a_{k+1}g^{k+1}
$$
for fixed $N$ and small $g$.

In general, the nature of an asymptotic series is such that the
correct sum is uniformly approached until, after an optimum number
$N_0$ of terms the subsequent terms drive the partial sums away from
the correct result for the total sum, leading to divergence
\cite{Hardy48}.

For models with a simple $O(m)$--symmetry the factorial growth of the
RG functions coefficients for large orders of perturbation theory is
well established \cite{Lipatov77}:
 \begin{equation}
 \label{assymptotics}
a_k = k! (-a)^k k^b c [1+O(\frac{1}{k})], \,\,\,\,\,
k\rightarrow\infty \end{equation}
the quantities $a$ and $b$ have been calculated in
Ref. \cite{Brezin77}, the values of constant $c$ in Ref.
\cite{Brezin78}. The estimate (\ref{assymptotics}) is a fundamental
one  since it shows that the initial functions can be reconstituted from
their asymptotic expansions by the application of the integral Borel
transformation \cite{borelsummability} in different modifications.
As it was already noted above the asymptotic nature of the divergence
of the RG  functions for the disordered models still has not been proved.

Below we will describe the methods of resummation which in the
field--theore\-ti\-cal RG scheme are applied to study the critical
behavior of the Stanley model (\ref{phi4}) as well as the disordered
models (\ref{dil_phi4}), (\ref{hamlr}), (\ref{saw}), (\ref{hamis}),
(\ref{hamcub}).

\subsection{Resummation of asymptotic series with one
coupling}\label{VB}

The idea of resummation which can be
applied to an asymptotic series consists in changing the order of
summation\cite{Hardy48}. In the case of one variable it
means that starting from a power series
\begin{equation}
S\left( x\right) =\lim _{L\rightarrow \infty }\sum_{i=1}^L a_ix^i,
\end{equation}
one performs an identical transformation \cite{footnote4}
\begin{eqnarray}
&S\left( x\right) \equiv \lim _{L\rightarrow \infty
}\sum_{i=1}^L(\frac{ a_ix^i}{i!}\int_0^\infty dt\exp (-t)t^i)\equiv
\nonumber&\\ &\lim _{L\rightarrow \infty}\{\lim _{A\rightarrow \infty
}\int_0^Adt [\sum_{i=1}^L\frac{a_i(xt)^i}{i!}\exp (-t)]\},
\end{eqnarray}
and redefines the sum by the following expression:
\begin{eqnarray}
&S^{\prime }\left( x\right) =\lim _{A\rightarrow \infty }\{\lim
_{L\rightarrow \infty }\int_0^Adt[\sum_{i=1}^L\frac{a_i(xt)^i}{i!}\exp
(-t)]\}\equiv \nonumber&\\ &\int_0^\infty dt\exp (-t)\sum_{i=1}^\infty
\frac{a_i(xt)^i}{i!},
\end{eqnarray}
where $\sum_{i=1}^L a_i(xt)^i/i!$ is called the Borel - image of
$\sum_{i=1}^La_ix^i$. This trick is natural in the sense that
for the case of a convergent series one finds $S^{\prime }=S$ within the
radius of convergence.

The above mentioned procedure in the case of a single variable is
known as the Borel resummation technique and in different
modifications it is widely used for evaluating asymptotic series.
Unfortunately this technique cannot be applied to our case immediately
because only the partial sum of the truncated series is known. To overcome
this obstacle one represents the Borel-image of the initial sum in the
form of a rational approximant and in this way reconstitutes the
general form of the series. This technique that involves a rational
approximation and the Borel transformation is known as the
Pad\'e-Borel resummation technique \cite{Baker} and is performed as
follows:
\begin{itemize}
\item
the Borel-image of the initial sum is constructed:
\begin{equation}
\label{BorIm}
\sum_{i=1}^La_ix^i\Rightarrow \sum_{i=1}^L\frac{a_i(xt)^i}{i!};
\end{equation}
\item
the Borel-image is extrapolated by a rational approximant
\begin{equation}\label{Ratapr}
\left[ M/N \right] =\left[ M/N \right] (xt);
\end{equation}
here, $\left[ M/N \right]$ stands for the quotient of two
polynomials in $xt$; $M$ is the order of the numerator and $N$
is that of the denominator;
\item
the resummed function is then obtained by the integral:
\begin{equation}\label{res}
S^{res}(x)=\int_0^\infty dt \exp (-t)\left[ M/N \right] (xt).
\end{equation}
\end{itemize}
The Pad\'e-Borel procedure can be generalized by introducing an
additional fit parameter $p$ to the Borel transformation.
Substituting the factorial $i!$ by the Euler gamma-function
$\Gamma(i+p+1)$ and inserting an additional factor $t^p$ into the
integral (\ref{res}), one defines the Pad\'e-Borel-Leroy resummation
procedure.

The Pad\'e-Borel as well as Pad\'e-Borel-Leroy resummation
procedures are applicable to RG functions of one coupling constant
such as those of model (\ref{phi4}). For models with more
complicated symmetries and several couplings one uses the
generalizations of the Pad\'e--Borel method as explained in
the following subsection.

\subsection{The case of several couplings}\label{VC}

In order to suit the resummation procedure (\ref{BorIm})-(\ref{res})
for functions that depend on several variables one should change
the first step (\ref{BorIm}): for example, for the two-variable case
one defines the Borel image by \cite{footnote5}:
\begin{equation}\label{BorIm1}
\sum_{i,j}a_{i,j}x^iy^j\Rightarrow
\sum_{i,j}\frac{a_{i,j}(xt)^i(yt)^j}{(i+j)!}.
\end{equation}
The rational approximation (\ref{Ratapr}) can be performed then
either in the dummy variable $t$ or in the variables $x,\,y$. This
defines two approaches that one can use to resum the functions of
several variables. In the first case the resummation procedure is
referred to as the Pad\'e-Borel resummation for resolvent series
\cite{Watson74}. The application of the Chisholm approximants
\cite{Chisholm73} which are the generalization of
Pad\'e--approximants to the many--variable case is necessary in
the second case. A Chisholm approximant can be defined as a ratio
of two polynomials both in variables $x$ and $y$, of degree $M$
and $N$ such that the first terms of its expansion are equal to
those of the function which is approximated. Again, the
resummation is performed in eq. (\ref{res}) replacing the Pad\'e
with the Chisholm approximant. This method will be referred below
as the Chisholm-Borel resummation.

It is obvious that an initial sum can be resummed in different
ways. Apart from the  Leroy fit parameter $p$ mentioned above some
arbitrariness arises from the different types of rational
approximants one may construct. For instance, within the two-loop
approximation the method of Pad\'e-Borel resummation of a
resolvent series can be done using either the $\left[ 0/2 \right]$
or the $\left[ 1/1 \right]$ approximants. The Chisholm-Borel
approximation implies even more arbitrariness and demands a very
precise analysis of the approximants to be chosen.

We want to stress here again that in contrast to the single variable
case the validity of neither the Pad\'e-Borel nor the Chisholm-Borel
resummations have been proven rigorously for any cases with two or
more variables. The success of their application can solely be judged
by the consistency of the results with both experimental and
Monte-Carlo data.

\subsection{The method of subsequent resummation}\label{VD}

In the resummation schemes described above for series in several
variables the variables are treated as ``equal in rights".
This is not the case in the recently proposed  method of
so-called subsequent resummation, developed in the context of
the $d=0$-dimensional disordered Ising model in Ref. \cite{Alvarez00}.

Starting from the RG function $f(u,v)$ in the form of a series in
powers of $u,v$ (for the model (\ref{dil_phi4}), $u$ is the original
coupling of the undiluted system and $v$ is the variance of the
quenched disorder)
\begin{equation}
f(u,v)=\sum_{n}\sum_{k}C_{k,n}u^k v^n,
\label{s}
\end{equation}
one rewrites it as a series in the variable $v$:
\begin{equation}
f(u,v)=\sum_{n}A_n(u) v^n,
\label{ss}
\end{equation}
with coefficients that are in turn series in the variable $u$ and are to
be resummed in advance by any one-variable method:
\begin{equation}
A_n(u)\equiv\sum_{k}C_{k,n}u^k.
\label{sss}
\end{equation}
The main result of Ref. \cite{Alvarez00} is that the expansions of the
coefficients (\ref{sss}) and the resulting series (\ref{ss}) at fixed
$u$ are Borel summable (for a $d=0$ dimensional system, however). This
gives a hint to analyze the RG expansions of the $d=3$ disordered
models by first performing a Pad\'e-Borel resummations of the
corresponding series for the coefficients in one coupling and then,
using the computed coefficients, resumming the series in the other
subsequent coupling(s).

\section{Results of the renormalization group analysis}\label{VI}

Now, having at hand the powerful methods of the field-theoretical
renormalization group refined by resummation of the (asymptotic)
series we can check to what results does it lead when applied to
systems with weak quenched disorder. Again, we start from the
uncorrelated point-like disorder, then pass to the systems with
long-range correlated disorder, and finally we analyze
random-anisotropy systems.

\subsection{Weakly diluted Stanley model with point--like
uncorrelated defects}\label{VIA}

Referring to the universal critical properties of the model
(\ref{dil_phi4}) at $d=3$, as implied by the Harris criterion one
observes new critical behavior only for the case $m=1$, i.e. for the
weakly diluted quenched Ising model.

Following the procedure sketched in the subsection~\ref{IVD}
the expressions for the RG functions of the model are
obtained as series in the renormalized couplings $u$ and $v$. Each order
of the perturbation theory corresponds to a given number of integrations
in momentum space and, equivalently, to the number of loops in Feynman
diagrams in the diagrammatic representation for the vertex functions
(\ref{vertex}).

The massive RG functions of the weakly diluted quenched Ising model
are known to high orders of the perturbation theory. They result from
more than twenty years of laborious studies. Within the three--loop
accuracy the functions were first reported in Ref.  \cite{Sokolov77}.
They yet contained errors which were corrected in Ref.
\cite{Sokolov81}, erroneously again. The final three-loop expressions were
only published in Ref. \cite{Shpot89}.  Subsequently, four-loop series were
obtained in Ref. \cite{Mayer89}.  Recently five--loop \cite{Pakhnin00}
and record six--loop \cite{Carmona00} expansions became available.
Written in the two-loop approximation the functions read
\cite{Holovatch92}:
\begin{eqnarray}
\beta_u(u,v) & = & -(4-d) u \Big \{ 1 - u - \frac{3}{2} v +
\frac{8}{27} \big [ 9 (i_1-\frac{1}{2})+i_2 \big ] u^2 +\nonumber\\
&&\frac{2}{3} \big [ 12 (i_1-\frac{1}{2})+i_2 \big ] uv +
\frac{1}{4} \big [ 21 (i_1-\frac{1}{2})+i_2 \big ] v^2 \Big \},
\label{2lmass1}
\\
\beta_v(u,v) & = & -(4-d) v \Big \{ 1 - v - \frac{2}{3} u +
\frac{1}{4} \big [ 11 (i_1-\frac{1}{2})+i_2 \big ] v^2 +\nonumber\\
&&\frac{2}{3} \big [ 6 (i_1-\frac{1}{2})+i_2 \big ] vu +
\frac{8}{27} \big [ 3 (i_1-\frac{1}{2})+i_2 \big ] u^2 \Big \},
\label{2lmass2}\\
\gamma_{\phi}(u,v) & = & - 2 (4-d) \Big \{ \Big [
\frac{2}{27} u^{2} + \frac{1}{6} uv + \frac{1}{16} v^{2}\Big ] i_2
\Big \}, \label{2lmass3}
\\
\bar{\gamma}_{\phi^2}(u,v) & = &  (4-d) \Big \{
\frac{1}{3} u + \frac{1}{4} v -
12 \Big [ \frac{1}{27} u^2 +
\frac{1}{12} uv +
\frac{1}{32} v^2
\Big ] (i_1-\frac{1}{2}) \Big \}.
\label{2lmass4}
\end{eqnarray}
Here, we use the conventional normalization for the couplings $u$, $v$
such that the coefficients of $u$ and $v$ in the one-loop contribution to
$\beta_u$ and $\beta_v$ equal unity.

The expressions (\ref{2lmass1})--(\ref{2lmass4})
besides the couplings $u,\,v$ depend on space dimension $d$.
In particular, $d$ enters the two--loop integrals $i_1(d)$, $i_2(d)$.
The two--loop contributions to the RG functions of the
three-dimensional weakly diluted quenched Ising model can be
obtained from the expressions (\ref{2lmass1})--(\ref{2lmass4}) by
substituting $i_1(3)=2/3$, $i_2(3)=-2/27$ \cite{loopint}.

The explicit dependence of the functions
(\ref{2lmass1})--(\ref{2lmass4}) on the space
dimension $d$ was exploited in Ref. \cite{Holovatch92}.
Expanding the values of the loop-integrals $i_1,\,i_2$ in
$\varepsilon=4-d$, the functions were obtained in
the form of an $\varepsilon$--expansion \cite{Grinstein76}.
Classically, this
result follows the familiar $\varepsilon$-expansion
\cite{Wilson72} of the critical exponents. In order to obtain the
expansions, one solves the fixed point equations (\ref{fp}) with
$\varepsilon$ as a small parameter and then re-expands in
$\varepsilon$ the critical exponents after substitution
of the coordinates $u^*(\varepsilon)$, $v^*(\varepsilon)$ into the
$\gamma$-functions.

This form of analysis of the six--loop massive RG functions
encounters hardships since the $\varepsilon$-expansions of the loop
integrals are not known in high four-, five- and six--loop
approximations. This obstacle can be overcome by applying
an alternative dimensional regularization scheme with minimal
subtraction \cite{tHooft72}. In this approach the RG functions
were obtained in the three-loop \cite{Janssen95} approximation and
can be reconstituted within the five-loop accuracy from the RG
functions of the cubic model \cite{Kleinert95}. The degeneracy of
the one--loop RG functions lead to the result, that instead of
expanding in $\varepsilon$ one is to expand in
$\sqrt{\varepsilon}$ \cite{Grinstein76,sqrt}. This results in
$\sqrt{\varepsilon}$ expansions for the critical exponents
\cite{Shalaev97} and the stability matrix (\ref{stmatrix})
eigenvalues \cite{Folk99}:
\begin{eqnarray} \label{47}
\nu  &=& 0.5{+}
0.08411582\varepsilon^{1/2} { -} 0.01663203\varepsilon  {+}
0.04775351\varepsilon
^{3/2} {+} 0.27258431\varepsilon^2\,,
\\ \label{48}
\eta  &=&  {-}
0.00943396\varepsilon  {+} 0.03494350\varepsilon ^{3/2} {-}
0.04486498\varepsilon
^{2} {+} 0.02157321\varepsilon ^{5/2}\,,
\\ \label{49}
\gamma  &=& 1 {+}
0.16823164\varepsilon^{1/2} {-} 0.02854708\varepsilon  {+}
0.07882881\varepsilon
^{3/2} {+} 0.56450490\varepsilon ^{2}\,,
\label{sqrteps} \\ \label{50}
\omega_1&=&2\,
\varepsilon{+}3.70401119 \, \varepsilon^{3/2}{+}11.30873837 \,
\varepsilon^2 \, ,
\\ \label{51}
\omega_2&{=}&0.67292659\, \varepsilon^{1/2}{-}1.92550909 \,
\varepsilon
{-}0.57252518 \, \varepsilon^{3/2} {-}13.93125952 \, \varepsilon^2 .
\end{eqnarray}
In the above expressions the two--loop results for the exponents were
obtained in Ref.~\cite{Grinstein76}, the three--loop
results were presented independently in Refs. \cite{Shalaev77}
and \cite{Jayaprakash77}.

The crucial property of the $\sqrt{\varepsilon}$-expansions
(\ref{47})-(\ref{51}) is their formal character.
Though they allow to predict qualitatively new critical behavior in
the weakly diluted Ising model \cite{Grinstein76,sqrt} they seem to be
of no use for any quantitative analysis. Naively adding the
successive perturbational contributions in the
$\sqrt{\varepsilon}$-expansion for the model stability matrix
eigenvalues (\ref{50}), (\ref{51}) one observes that already in the
three--loop approximation ($\sim \varepsilon$) $\omega_2$ becomes
negative and therefore no stable fixed point exists in strict
$\sqrt{\varepsilon}$-expansion \cite{Folk99}. Even resummation
procedures do not change this picture
\cite{Folk00,Folk98}.

The way to cope with the bad convergence properties of the
$\sqrt{\varepsilon}$-expansion can be found in the fixed-dimension
approach which does not imply expanding the loop integrals in
$\varepsilon$.
This method yields most numerical results for the universal
characteristics of the diluted Ising model at criticality.
The numerical solution of the fixed point
equations (\ref{fp}) the resummed $\beta$-functions
(\ref{2lmass1}) and the resummation of the $\gamma$--functions
(\ref{2lmass4}) in the stable fixed point shows the validity of
the method. The study of the massive $\beta$-functions of the
diluted Ising model resummed in this way revealed
that starting from the two--loop approximation the random fixed
point is stable and is present
in all orders
of perturbation theory \cite{Shpot89,Mayer89,%
Pakhnin00,Holovatch92,%
Jug83,Mayer84,Mayer89a,%
Varnashev00,Holovatch98,Folk00a}.
Based on the three-dimensional values of the loop-integrals
\cite{loopint}, the RG functions of the model are calculated
\cite{Carmona00} and resummed \cite{Pelissetto} at present in the
six-loop approximation.

\begin{figure}[t]
\begin{center}
\includegraphics[width= 100mm]{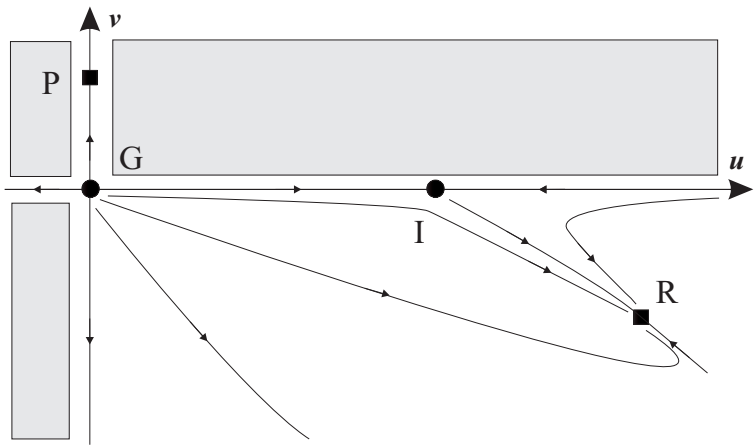}
\end{center}
\caption{Qualitative RG flow for the weakly diluted quenched Ising
model.  The Gaussian fixed point {\bf G} is stable for $d\geq4$,
stable fixed point {\bf P} can not be reached from the initial
coupling values $u>0$, $v<0$. The unphysical regions for the
random Ising model are shown in grey tone. The pure Ising fixed
point {\bf I} is unstable whereas random fixed point {\bf R} is
both stable and accessible (stable fixed points are shown by
boxes). \label{figfp} }
\end{figure}

Let us analyze the fixed point structure of the $\beta$-functions for
the random Ising model at $d=3$. Within the one--loop approximation
(i.e. including only linear terms in $u$, $v$ in the curly brackets in
(\ref{2lmass1}), (\ref{2lmass2})) at $d=3$ one finds three fixed
points (see Fig. \ref{figfp}): the Gaussian fixed point {\bf G}
$u^*=v^*=0$, the pure Ising fixed point {\bf I} $u^*\neq0$, $v^*=0$
and the ``polymer" fixed point {\bf P} $u^*=0$, $v^*\neq0$. It is
straightforward to check that the fixed points {\bf G} and {\bf I} are
unstable whereas the fixed point {\bf P} is stable but as far as
$v^*>0$ it is inaccessible for the initial values of couplings of our
model. Within the one-loop approximation one does not encounter the
fixed point {\bf R} with both non-zero coordinates $u^*\neq0$,
$v^*\neq0$: this happens because the system of equations for the fixed
points is degenerate on the one--loop level \cite{Grinstein76,sqrt}.
As we will see, this fixed point appears in the next, two--loop
approximation if the resummation is applied.

\begin{figure}[htbp]
\begin{center}
\includegraphics[width=100mm,height=70mm]{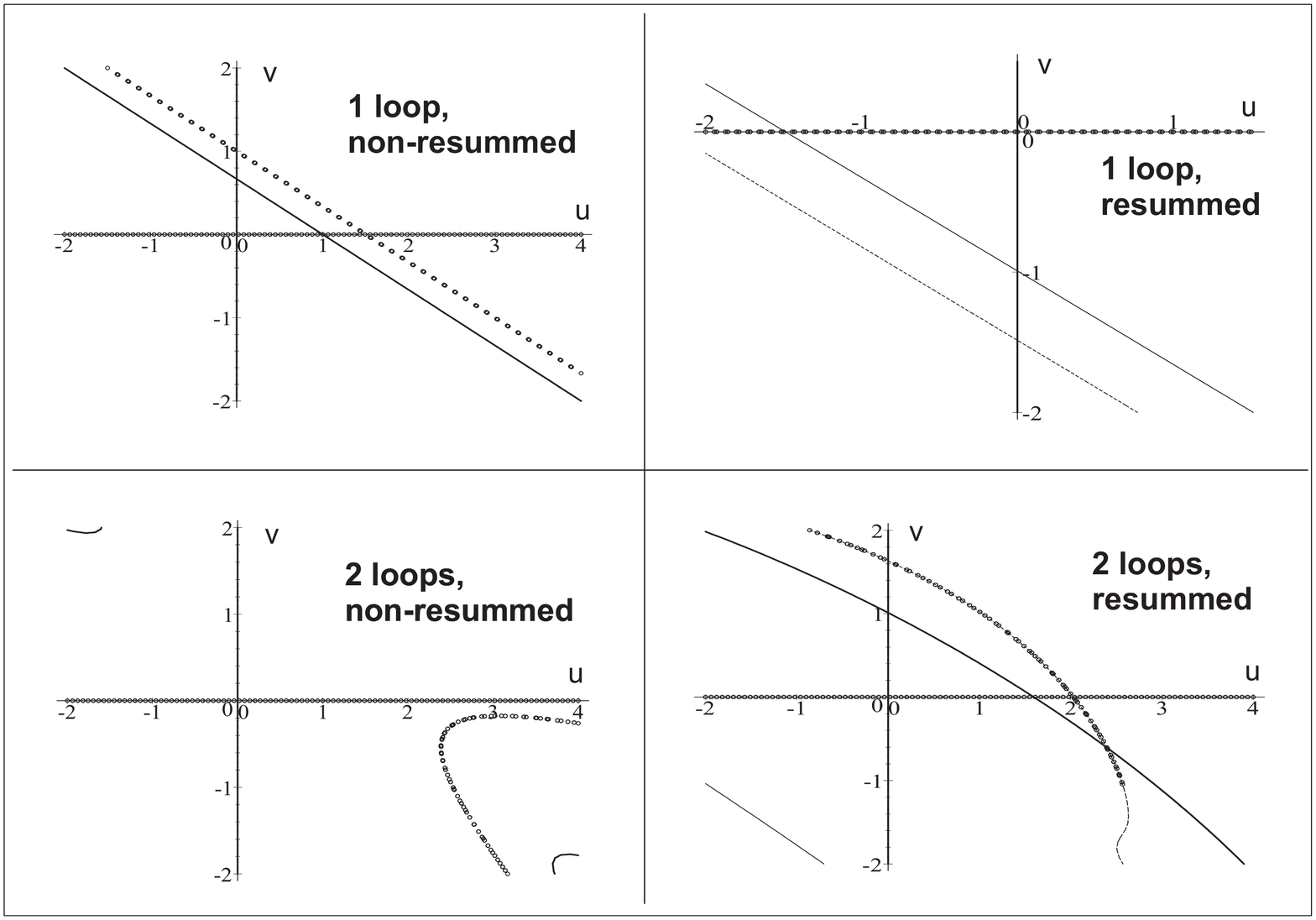}
\end{center}
\caption{\label{comp3} The lines of zeros of the diluted Ising
model massive
  $\beta$-functions in the one- and two-loop approximations of the
  perturbation theory, left-hand column: non-resummed, right-hand
  column: resummed by the Chi\-sholm-Borel method. Circles correspond
  to $\beta_u=0$, thick lines depict $\beta_v=0$. Thin solid and
  dashed lines show the zeros of the analytically continued functions
  $\beta_u$ and $\beta_v$ respectively. One can see the appearance of
  the mixed fixed point $u>0, v<0$ in the two--loop approximation for
  the resummed $\beta$-functions. The figure is taken from Ref.
  \protect\cite{Holovatch98}.  }
\end{figure}

\begin{figure}[t]
\begin{center}
\includegraphics[width=100mm,height=70mm]{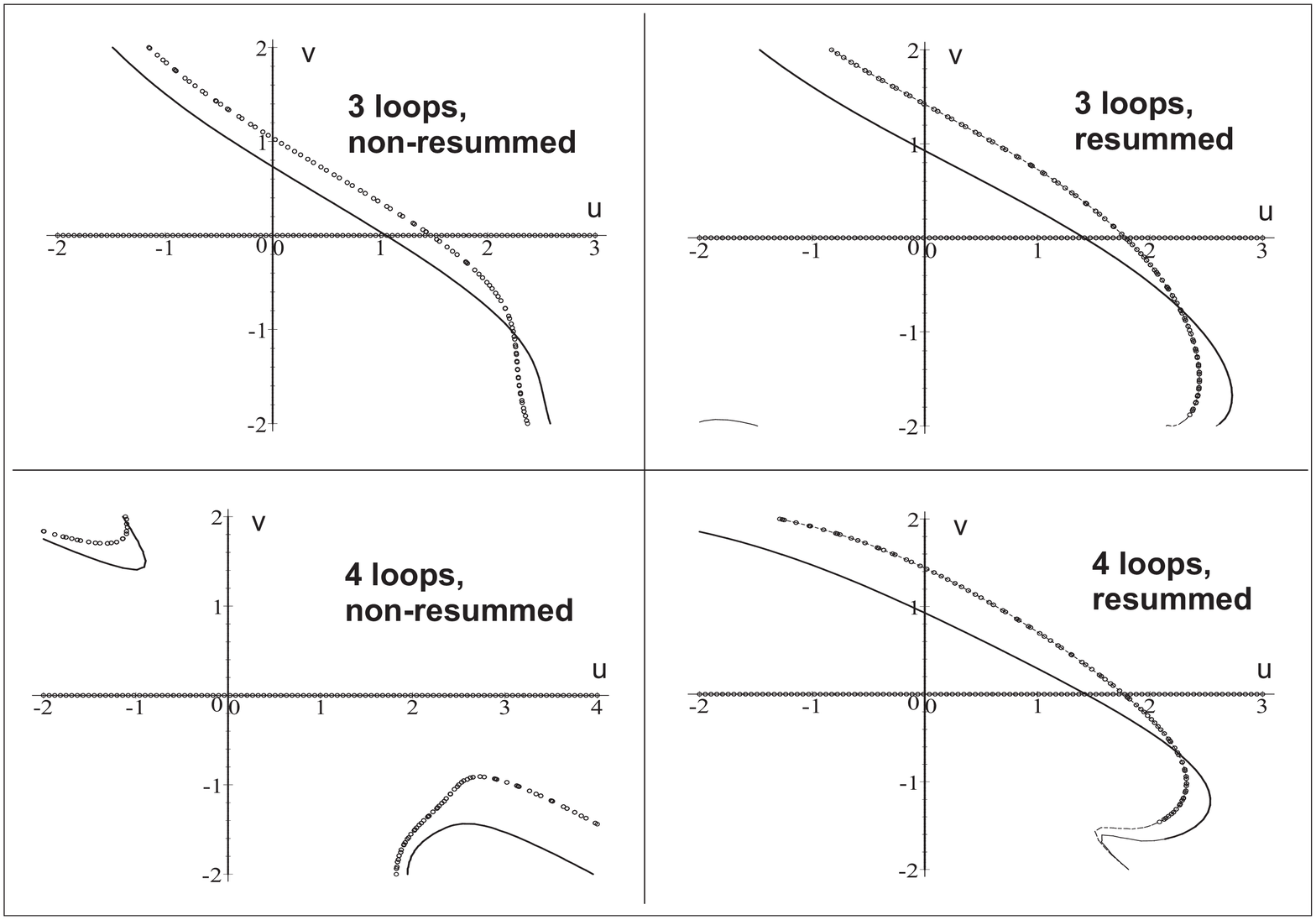}
\end{center}
\caption{\label{comp4} The lines of zeros of the diluted Ising
model massive
  $\beta$-functions in the three- and four-loop approximations of the
  perturbation theory, left-hand column: non-resummed, right-hand
  column: resummed by the Chi\-sholm-Borel method.  The notations are
  the same as in the figure \protect{\ref{comp3}}. Close to the mixed
  fixed point the behaviour of the resummed functions remains alike
  with the increasing order of approximation. This is not the
  case for the non-resummed functions. The figure is taken from Ref.
  \protect\cite{Holovatch98}.  }
\end{figure}

Applying the Chisholm-Borel resummation procedure of the
subsection \ref{VC} one encounters the random fixed point {\bf R}
of the model in the two-loop approximation (see Fig. \ref{figfp}).
The stability analysis shows that this fixed point is stable
proving the crossover to a new critical regime under dilution. In
Figs. \ref{comp3}, \ref{comp4} we show the curves
$\beta_u(u,v)=0$, $\beta_v(u,v)=0$ in the $u-v$ plane. The
intersections of these curves (i.e.  simultaneous zeros of both
$\beta$--functions) correspond to the fixed points. The ``naive''
analysis of the $\beta$--functions, without applying any
resummation procedure leads to the curves, which are shown on the
left-hand side of the Figs. \ref{comp3}, \ref{comp4}. Without
resummation only in the three--loop approximation one finds a
stable random fixed point $u^*\neq0, v^*\neq0$. However the fixed
point then disappears in the four--loop approximation. A
completely different picture is observed when the resummation
procedure is applied (right hand columns of the figures). In the
region of interest for the values of the couplings the topology of
the fixed point picture remains stable for increasing orders of
the approximation from the two--loop to the four--loop level.

The Figs.~\ref{comp3}, \ref{comp4} show also the historical
evolution of the study of the weakly diluted quenched Ising model.
Already in the two--loop level the RG functions resummed by the
Chisholm-Borel procedure \cite{Jug83} revealed that no
difficulties connected to the degeneracy of the $\beta$--functions
are encountered. Critical exponents extracted from the resummed
$\gamma$--functions values in the fixed point were found to be
clearly larger than those of the pure model (see Table
\ref{table_theory}). As noted above,  in the three--loop level the
straightforward analysis of the $\beta$--functions
\cite{Sokolov77,Sokolov81} yields fixed point coordinates and
critical exponents without resummation, but the accuracy obtained
did not allow to estimate, for instance the heat capacity critical
behavior. Numerical values of the random Ising model critical
exponents as obtained by the RG approach are given in Table
\ref{table_theory} with reference to the renormalization schemes
and resummation procedures \cite{errbars}. We note here a
breakthrough that occurred during several months of 1999-2000 when
the perturbation theory expansions were extended from the 4th
\cite{Mayer89} through 5th \cite{Pakhnin00} to the sixth order
\cite{Carmona00,Pelissetto}.

{\footnotesize
\begin{table}
\ttbl{31pc}{\label{table_theory}
  The theoretical values for the asymptotic critical exponents of the
  weakly diluted Ising model.  {\it n}LA denotes the {\it n}th order
  in loopwise approximation within the massive (`mass') and $3d$
  minimal subtraction (`MS') schemes of the field-theoretical
  renormalization group approach.  The resummation procedures are
  given in the following notations: ChBr -- Chisholm--Borel; PdBr --
  Pad\'e--Borel; AW -- $\varepsilon$ algorithm of Wynn, CM -- Borel
  transformation with conformal mapping.  SF stands for Golner-Riedel
  scaling field method, EEA denotes nonperturbative RG approach based
  on the concept of effective average action, superscript $^c$ at the
  correction-to-scaling exponent $\omega$ indicates that the real part
  of the corresponding complex number is shown.  \newline}
{\begin{tabular}{|c|c|c|c|c|c|c|c|}
\hline
Ref. & RG  &
Order & Resum- & $\nu$ & $\eta$ & $\gamma$ & $\omega$
\\
& scheme & &mation&&&&\\
\hline
Sokolov et al.,& mass & 3LA & No & & 0.009 & 1.31 &
\\1981, Ref. \cite{Sokolov81}&&&&&&& \\
\hline Newman et al.,& SF & No &  & 0.70 & 0.015 & 1.39 &
0.41
\\1982, Ref. \cite{Newman82}&&&&&&& \\
\hline Jug, 1983, & mass & 2LA & ChBr & 0.678 & 0.031 & 1.336 &
0.450$^{c}$
\\  Ref. \cite{Jug83}&&&&&&& \\
\hline
Mayer et al.,& mass & 2LA & ChBr & & 0.031 & 1.337 &
\\
1984, Ref. \cite{Mayer84}&      & 3LA & ChBr & & 0.022 & 1.325 &
\\
\hline
Mayer et al.,& mass & 4LA & ChBr & 0.670 & 0.034 & 1.326 &
\\1989,  Ref. \cite{Mayer89}&&&&&&& \\
\hline Mayer, 1989, & mass & 4LA & AW & 0.6680 & & 1.318 &
\\
Ref. \cite{Mayer89a}&      & 4LA & PdBr & 0.6714 & & 1.321 &
\\
\hline Shpot, 1989, & mass & 3LA & ChBr & 0.671 & 0.021 & 1.328 &
0.359
\\  Ref. \cite{Shpot89}&&&&&&& \\
\hline
Janssen et al.,& MS, 3d & 3LA & PdBr & 0.666 & & 1.313 & 0.366
\\1995,  Ref. \cite{Janssen95}&&&&&&& \\
\hline
Holovatch et al.,& mass & 3LA & ChBr & 0.671 & 0.019 &
 1.328 & 0.376
\\1997,  Ref. \cite{Holovatch98} &&&&&&& \\
\hline
    &        & 2LA &           & 0.665 & 0.032 & 1.308 &
0.162
\\
Folk et al.,& MS, 3d & 3LA & ChBr  & 0.654 & 0.022 & 1.293 &
0.430
\\
1998, Ref. \cite{Folk98}&        &  4LA  &  & 0.675 & 0.049 &
1.318 & 0.390$^c$
\\
\hline Folk et al.,& MS, 3d & 4LA & ChBr & & & &
0.39(4)$^{c}$
\\1999,  Ref. \cite{Folk99}& mass & 4LA & ChBr & & & & 0.372(5)
\\
\hline
Pakhnin et al.,& mass & 5LA & PdBr & 0.671(5) & 0.025(10)&
1.325(3) &
0.32(6)
\\2000,  Ref. \cite{Pakhnin00}&&&&&&& \\ \hline
Varnashev,& mass & 4LA & PdBr & 0.681(12) & 0.040(11) &
 1.336(20) &
\\
2000 Ref. \cite{Varnashev00}&      &     & PdBr & 0.672(4) & 0.034(10)
& 1.323(10) &
0.330
\\
\hline Pelissetto et al.,& mass & 6LA & PdBr- & 0.678(10) &
0.030(3) & 1.330(17) & 0.25(10) \\
&&&CM&&&&\\
2000, Ref. \cite{Pelissetto}& &
& PdBr-& 0.668(6)  & 0.0327(19)  &  1.313(14)  & 0.25(10)\\
&&&PdBr&&&&\\
\hline Tissier et al.,& EEA & No &  & 0.67 & 0.05 & 1.306 &
 \\
2001, Ref. \cite{Tissier01} & & & &   & &  &\\ \hline
\end{tabular}}
\end{table}
}

Already the 5th loop order Pad\'e-Borel resummation meets some
difficulties. Moreover, the extension to the six--loop order revealed
a wide gap between the five-- and six--loop fixed point coordinates
\cite{Folk00a} and subsequently an inconsistency of the six--loop
values of critical exponents compared with the five--loop results of
Refs \cite{Pakhnin00}.  One also encounters a similar behavior of the
resummed series when the $d=3$ minimal subtraction RG technique is
applied to the random Ising model: the random fixed point is present
in the resummed two- three- and four- loop approximations whereas it
disappears in the five-loop approximation \cite{Folk00,Folk98}.  This
lead to the conjecture about the possible Borel-non-summability of the
series.

The fact that the weak coupling expansion might not be of
asymptotic nature for the random Ising model was predicted already
in a study of the randomly diluted model in zero dimensions
\cite{nonsum}. The perturbation theory series of this toy-model
appeared to be Borel non-summable. Moreover, this property was
shown to be a direct consequence of the appearance of
Griffiths-like singularities \cite{Griffiths69} that are caused by
the zeroes of the partition function of the pure system.

However, as noted above, in another recent study of the $d=0$
dimensional random Ising model it was shown analytically
\cite{Alvarez00} that the perturbative expansions for the free
energy are Borel summable provided the summation is carried out
subsequently, as described above in subsection \ref{VD}. The
application of this procedure to the case $d=3$ reconstituted
\cite{Pelissetto} earlier data for the asymptotic critical
exponents of the model in massive scheme and allowed to restore
the presence of the random fixed point in the $d=3$ minimal
subtraction scheme on the five loop level \cite{Blavats'ka01b}.

Alternatively, the $d=3$ random Ising model was treated
non-perturbatively first by means of the Golner-Riedel scaling
field approach \cite{Newman82}, and very recently by an RG
approach based on the concept of effective average action
\cite{Tissier01}.

\subsection{Systems with long-range-correlated disorder}\label{VIB}

\subsubsection{Magnets with long-range correlated disorder}\label{VIB1}

Now, let us discuss the results obtained for the $m$-vector model with
long-range-correlated disorder described by the effective Hamiltonian
(\ref{hamlr}).  A one-loop approximation was given in Ref.
\cite{Weinrib} using an expansion in $\varepsilon=4-d,\delta=4-a$. A
new long-range-correlated fixed point was found with a
correlation-length exponent $\nu=2/a$ and it was argued, that this new
scaling relation is exact and also holds in higher order
approximations.  This result was questioned recently in Refs.
\cite{Prudnikov}, where the static and dynamic properties of 3d
systems with long-range-correlated disorder were studied by means of
the massive field-theoretical RG approach in a 2-loop approximation,
for different fixed values of the correlation parameter, $2\leq a\leq
3$.  The $\beta$ and $\gamma$ functions in the two-loop approximation
were calculated as an expansion series in renormalized vertices $u$,
$v$ and $w$. The Pad\'e-Borel resummation was used for their analysis.
The study revealed the existence of the stable random fixed point with
$u^* \neq 0, \, v^* \neq 0, \, w^* \neq 0$ in the whole region of the
parameter $a$. The comparison of the obtained exponents $\nu$ values
at various $a$ and ratio $2/a$ shows the violation of the supposed
 exact the relation $\nu = 2/a$ of ref.\cite{Weinrib}.  The authors
concluded that the revealed difference is caused by the use of a more
accurate field-theoretical description using higher order
approximations for the 3d system directly together with methods of
series summation. Recently, the correction-to-scaling exponent for
magnets with long-range-correlated defects was obtained using
the same approach \cite{Blavats'ka01a}.

\subsubsection{Polymers in media with long-range-correlated quenched
disorder} \label{VIB2}

The effective Hamiltonian (\ref{saw}) is the starting point to
study the polymer limit $m\rightarrow0$ of the weakly diluted
Stanley model with long-range-correlated disorder. As we explained
in the subsection \ref{IIIE}  it may serve as a model for
polymers in porous media. Again, imposing the renormalization
conditions of the massive scheme (\ref{conditions}),
(\ref{adcond}) the one-loop approximation leads to the
following expressions for the RG functions \cite{Blavats'ka01}:
\begin{equation}
\beta_u = - \varepsilon \left[ u- \frac{4}{3}\,u^2
I_1 \right ]-\delta 2uw \left[ I_2+\frac{1}{3} I_4 \right] + (2
 \delta- \varepsilon) \frac{2}{3}\,w^2 I_3 , \label{A}
\end{equation}
\begin{equation} \label{A1}
\beta_w = -\delta \left [w +
\frac{2}{3}\,w^2 I_2  \right ] +\varepsilon  \frac{2}{3} \left[wu I_1-
 w^2 I_4 \right] ,
 \end{equation}
 \begin{equation} \bar{\gamma}_{\phi^2}=
\varepsilon \frac{u}{3}\, I_1 -\delta \,\frac{w}{3}\,I_2 ,
\phantom{55} \gamma_{\phi}=\delta\, \frac{w}{3} I_4.  \label{B}
 \end{equation}
Here, $I_i$ are one-loop integrals that depend on the
space dimension $d$ and the correlation parameter $a$:
\begin{eqnarray} \label{int} I_1= \int\frac{ {\rm d}\vec{q}}{(q^2 +
1)^2 },\,\phantom{5} I_2&=& \int\frac{ {\rm d}\vec{q}\, q^{a-d}}{(q^2 +
1)^2 },\,\phantom{5} I_3= \int\frac{ {\rm d}\vec{q}\,
q^{2(a-d)}}{(q^2 + 1)^2 },\nonumber\\
&I_4=&\frac{\partial }{\partial
 k^2}\left [ \int\frac{ {\rm d}\vec{q}\, q^{a-d}}{[q+k]^2 + 1 }\right
]_{k^2=0}.
\end{eqnarray}
Note that in contrast to the usual $\phi^4$ theory the $\gamma_{\phi}$
function in Eq.~(\ref{B}) is nonzero already at the one-loop level.
This is due to the $k$-dependence of the integral $I_4$ in
Eq.~(\ref{int}).

\begin{figure}[t]
\begin{center}
\includegraphics{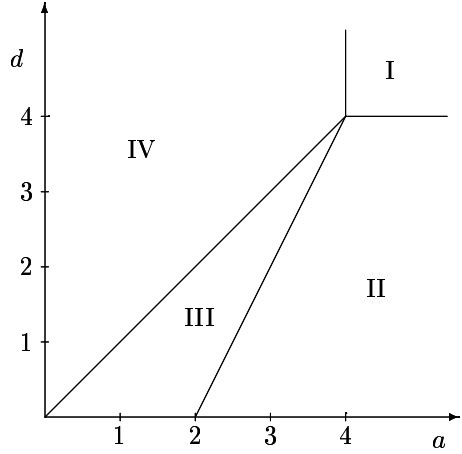}
\end{center}
\caption{\label{region}
  The critical behavior of a polymer in a medium with
  long-range-correlated disorder in different regions of the
  $d,a$-plane as predicted by the first order
  $\varepsilon,\delta$--expansion \protect\cite{Blavats'ka01}.  Region
  I corresponds to the Gaussian random walk behavior, in the region II
  scaling behavior is the same as in the medium without disorder, in
  region III the ``long-range'' fixed point {\bf LR} is stable and the
  scaling laws for polymers are altered, in region IV no accessible
  stable fixed points appear; this may be interpreted as the collapse
  of the chain.}
\end{figure}

There are essentially two ways to proceed in order to obtain the
qualitative characteristics of the critical behavior of the model.
The first is to consider the polynomials in Eqs. (\ref{A}),
(\ref{A1}) for fixed $a,d$ and look for the solution of the fixed
point equations. It is easy to check that these one-loop equations
do not have any stable accessible fixed points for $d<4$. The
second scheme is to evaluate these equations in a double expansion
in $\varepsilon=4-d$ and $\delta=4-a$ as proposed by Weinrib and
Halperin \cite{Weinrib}.  Substituting the loop integrals in
Eqs.~(\ref{A})-(\ref{B}) by their expansion in $\varepsilon=4-d$
and $\delta=4-a$ , one obtains the 3 fixed points given in the
table \ref{pol}.
\begin{table}[htp]
\ttbl{30pc}{\label{pol}Fixed points and stability matrix eigenvalues
in the first order of the $\varepsilon, \delta$~-~expansion
\protect\cite{Blavats'ka01}.\newline}
{\begin{tabular}{|c|cccc|}
\hline
Fixed Point & $u^{\ast}$ & $w^{\ast}$ & $\omega_1$ &
$\omega_2$\\ \hline Gaussian ({\bf G}) & $0$ & $0$ & $-\varepsilon$ &
$-\delta$\\ \hline Pure SAW ({\bf P}) & $\varepsilon$ & $0$ &
$\varepsilon$ & $\varepsilon/2-\delta$\\ \hline Long-range ({\bf LR})  &
$\frac{2\delta^2}{(\varepsilon-\delta)}$  &
 $-\frac{\delta(\varepsilon-2\delta)}{(\varepsilon-\delta)}$ &
$\phantom{55555555555}\frac{1}{2}\{\varepsilon-4\delta\pm\sqrt
{\varepsilon^2-4\varepsilon\delta+8\delta^2}\}$ &  \\
\hline
\end{tabular}}
\end{table}
The following conclusions may be drawn from these
first order results: Three distinct accessible fixed points are
found to be stable in different regions of the $a,d$-plane: the
Gaussian ({\bf G}) fixed point, the pure ({\bf P}) SAW fixed point
and the long-range ({\bf LR}) disorder SAW fixed point. The
corresponding regions in the $a,d$-plane are marked by I, II and
III in Fig.~\ref{region}. In the region IV no stable fixed point
is accessible.

For the correlation length critical exponent of the SAW, one finds
distinct values
$\nu_{\rm pure}$ for the pure fixed point and $\nu_{\rm LR}$ for the
long-range fixed point.  Taking into account
that the accessible values of the couplings are $u>0$, $w>0$, one
finds that the long-range stable fixed point is accessible only for
$\delta<\varepsilon<2\delta$, or $d<a<2+d/2$, a region where power
counting in Eq.~(\ref{saw}) shows that the disorder is irrelevant.
In this sense the region III for the stability of the {\bf LR} fixed point
is unphysical.
Formally, the first order results for $d<4$ read:
 \begin{equation}
 \label{9}
\nu = \left \{ \begin{array}{ll} \nu_{\rm pure}=
1/2 + \varepsilon/16,\; & \delta<\varepsilon/2, \\
\nu_{\rm LR}= 1/2 + \delta/8,\; &
\varepsilon/2 <\delta<\varepsilon.
\end{array}
\right.
\end{equation}
Thus, in this linear approximation the asymptotic behavior of
polymers is governed by a distinct exponent $\nu_{\rm LR}$ in the
region III of the parameter plane $a,d$. However, the region where the
{\bf LR} fixed point is found appears to be unphysical.

Something similar happens if the $\varepsilon,\delta$--expansion is
applied to study models of $m$--vector magnets with
long-range-correlated quenched disorder \cite{Weinrib,Blavats'ka01a}:
also in the case of magnets, as well as for polymers the first order
$\varepsilon,\delta$--expansion leads to a controversial phase
diagram. In order to obtain a clear picture and more reliable
information, one should proceed to higher order calculations.

To investigate the peculiarities of the critical behavior in the
2-loop approximation one may make use of the $m\to  0$ limit of
the appropriate $m$-vector model (\ref{hamlr}), investigated
recently \cite{Prudnikov}. Starting from the two-loop expressions
of Ref. \cite{Prudnikov} for the RG functions of the Stanley model
with long-range-correlated disorder and making use of the symmetry
arguments \cite{Kim,Blavats'ka01} in the polymer limit $m=0$  one
gets the following expressions for the $d=3$ RG functions of the
model described by the effective Hamiltonian (\ref{saw})
\cite{Blavats'ka01a}:

\begin{eqnarray}
\beta_u(u,w) & = & -u+u^2-(3f_1(a)-f_2(a))uw -
\frac{95}{216}u^3+\frac{1}{8}b_2(a)u^2w- \nonumber
\\
&&(b_3(a)-\frac{1}{4}b_6(a))uw^2 +f_3(a)w^2+b_5(a)w^3
\label{beta},
\end{eqnarray}
\begin{eqnarray}\label{b2}
\beta_w(u,w)&=&-(4-a)w-(f_1(a)-f_2(a))w^2+\frac{uw}{2}+b_{10}(a)
w^3- \nonumber \\
&&\frac{23}{216}u^2w+\frac{1}{4}b_{12}(a)uw^2,
\end{eqnarray}
\begin{equation}\label{b3}
\gamma_{\phi}(u,w)=\frac{1}{2}f_2(a)w+\frac{1}{108}u^2+c_1(a)w^2-
\frac{1}{4}c_2(a)uw,
\end{equation}
\begin{equation}\label{b4}
\bar{\gamma}_{\phi^2}(u,w)=\frac{1}{4}u-\frac{1}{2}f_1(a)w-\frac{1}{16}u^2-
c_3(a)w^2+ \frac{1}{4}c_4(a)uw.
\end{equation}
Here, the coefficients $f_i(a)$ are expressed in terms of the one-loop
integrals in Eq. (\ref{int}), $b_i(a)$ and $c_i(a)$
originate from the two-loop integrals and are tabulated in Ref.
\cite{Prudnikov} for $d=3$ and different values of the parameter $a$
in the range $2\leq a \leq 3 $. The series are normalized by a
standard change of variables
$u\to\frac{3u}{4}I_1,w\to\frac{w}{32}I_1$, so that the
coefficients of the terms $u,u^2$ in $ \beta_u $ become $1$ in
modulus.

\begin{figure}[bt]
\begin{center}
\includegraphics[height=50mm]{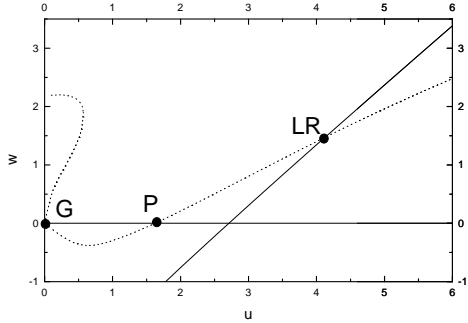}
\end{center}
\caption{ \protect{\label{fig5}} The lines of zeroes of the 3d
$\beta$-functions (\protect{\ref{beta}}), (\protect{\ref{b2}})
resummed by the Chisholm-Borel method at $a=2.9$. The dashed line
corresponds to $\beta_u=0$, the solid lines depict $\beta_w=0$.
The intersections of the dashed and solid lines give three fixed
points shown by filled circles at $u^{\ast}=0, w^{\ast}=0$ ({\bf
G}), $u^{\ast}=1.63, w^{\ast}=0$ ({\bf P}), and $u^{\ast}=4.13,
w^{\ast}=1.47$ ({\bf LR}). The fixed point {\bf LR} is stable. }
\end{figure}

As in the former cases discussed in subsections \ref{VIA},
\ref{VIB1} the question about the summability of the series in
Eqs.~(\ref{beta}) -- (\ref{b4}) is open. In Ref.
\cite{Blavats'ka01a} various kinds of resummation techniques have
been applied in order to obtain reliable quantitative results for
the system under consideration and to check the stability of these
results. First, a simple two-variable Chisholm-Borel resummation
technique was employed, which turns out to be the most effective
one for the given problem. In addition to the familiar fixed
points {\bf G} and {\bf P} describing Gaussian chains and polymers
(SAWs) on regular lattices, the stable long-range fixed point {\bf
LR} for polymers in long-range-correlated disorder was found.  In
Fig. \ref{fig5} the lines of zeroes of the resummed
$\beta$-functions (\ref{beta}), (\ref{b2}) at $a=2.9$ in the
$u,w$-plane in the region of interest are depicted.
The intersections of these curves correspond to the fixed points.
The corresponding values of the stable fixed point coordinates
and the stability matrix eigenvalues for different values of the
correlation parameter $a<3$ are given in Table \ref{chis}.

To calculate the critical exponents the same resummation technique
was applied. The numerical values for $\nu$, $\gamma$ and $\eta$
are listed in Table \ref{chis} for $a=2.3, \dots ,2.9$.  Note,
that for $a=3$, which corresponds to short-range-correlated
point-like defects, the interactions $u$ and $w$ become of the
same symmetry, so one can pass to one coupling $(u-w)$ and
reproduce the well-known values of the critical exponents for the
pure SAW model. The numerical values corresponding to those listed
in Table \ref{chis} in this case read: $u^*= 1.63$, $\nu= 0.59$,
$\gamma=1.17$, $\eta= 0.02$, $\omega= 0.64$. As departing from the
value $a=3$ downward to $2$ one notices a major increase of the
value of the coupling $u$, so the results are more reliable for
$a$ close to 3. At some value $a=a_{\rm marg}$ the {\bf LR} fixed
point becomes unstable.

\begin{table}
\ttbl{30pc}{\label{chis} Stable fixed point of the  3d 2-loop
$\beta$-functions, resummed by Chisholm-Borel method, the
corresponding critical exponents and the stability matrix
eigenvalues at various values of $a$. \newline}
{\begin{tabular}{|c|c|c|c|c|c|c|}
\hline $a$ &  $u^*$  & $w^*$ & $\nu$ & $\gamma$ & $\eta$ &
$\omega_{1,2}$ \\ \hline 2.9 & 4.13 & 1.47 & 0.64 & 1.25 & 0.04 &
0.25 $\pm$ 0.62 i \\ 2.8 & 4.73 & 1.68 & 0.64 & 1.26 & 0.04 & 0.22
$\pm$ 0.76 i \\ 2.7 & 5.31 & 1.81 & 0.65 & 1.28 & 0.03 & 0.18
$\pm$ 0.89 i \\ 2.6 & 5.89 & 1.87 & 0.66 & 1.29 & 0.03 & 0.15
$\pm$ 0.99 i \\ 2.5 & 6.48 & 1.89 & 0.66 & 1.31 & 0.02 & 0.11
$\pm$ 1.09 i \\ 2.4 & 7.10 & 1.87 & 0.67 & 1.33 & 0.01 & 0.07
$\pm$ 1.18 i \\ 2.3 & 7.76 & 1.84 & 0.68 & 1.36 & 0.01 & 0.03
$\pm$ 1.26 i \\ \hline
\end{tabular}}
\end{table}

To verify the results, the method of subsequent resummation (see
subsection \ref{VD}) was applied to the RG functions (\ref{beta}),
(\ref{b2}). Here, the summation was carried out first in the
coupling $u$ and subsequently in $w$. Again, the presence of
stable fixed point {\bf LR} for $a_{\rm marg} \leq a<d$ was
confirmed.

The obtained results may be summarized and interpreted  as follows:

(i) A new stable fixed point ({\bf LR}) for polymers in long-range-correlated
disorder is found for $d=3$, $a<d$, leading to critical exponents that
are different from those of the pure model;

(ii) There is a marginal value $a_{\rm marg}$ for the parameter $a$,
below which the stable fixed point is absent, indicating a chain
collapse of the polymer  for disorder that is stronger correlated.

(iii) The critical exponent $\nu$ increases with decreasing parameter
$a$, like in the Weinrib and Halperin case. But the
relation $\nu=2/a$ does not hold. Physically this means that in weak
long-range-correlated disorder ($a>a_{\rm marg}$) the polymer coil
swells with increasing correlation of the disorder. The self
avoiding path of the polymer has to take larger deviations to avoid
the defects of the medium.

\subsection{Critical behaviour of Stanley model with random
anisotropy}\label{VIC}

The RG results discussed so far in the subsections \ref{VIA},
\ref{VIB} concerned random-temperature like disorder. Another type
of disorder that we consider here is the disorder induced by a
single-ion anisotropy term in the RAM Hamiltonian (\ref{ram}).
According to the chosen distributions of the anisotropy axis
(\ref{dist1}) and (\ref{dist2}) our consideration falls into two
parts. First we analyze the case of the RAM with isotropic
distributions of anisotropy axis and then we discuss what
peculiarities in the critical behavior of the RAM might appear if the
random-anisotropy axis is taken to point only along the edges of
$m$-dimensional hypercube (we call this the "cubic" distribution).

\subsubsection{An isotropic distribution of the random anisotropy
axis}\label{VIC1}

Applying the renormalization conditions of the massive scheme
(\ref{conditions})  to the effective Hamiltonian (\ref{hamcub})
one obtains the  two-loop RG functions, that in the replica
limit $n=0$ read \cite{Dudka01a}:
\begin{eqnarray}
\label{betais1} {\beta_u}&=&-(4-d)\,u\Biggl
\{1-\frac{1}{6}\left[\left (m{+}8\right )u{+}12v{+}2 \left
(m{+}5\right )w\right]+ \frac{1}{9}\Bigl [2\left (5m{+}22\right
){u}^{2}{+}84{v}^{2}{+}\Bigr.\Biggr. \nonumber\\
&& 24 \left (m{+}5\right )uv{+}\Bigl.
 2\left (14m {+}58\right )wu{+}4\left
(9m{+}33\right )vw{+} \left (17m{+}67\right ){w}^{2}\Bigr ]
{i_1} +\frac{2}{9}\Bigl[ 2{v}^{2}{+}\Bigr. \nonumber
\\
&& \Biggl. \Bigl.2\left (m{+}2\right )uv{+}
\left (m{+}2\right ){ u}^{2}{+}\frac{m{+}3}{2}{w}^{2}{+}2\left
(m{+}2\right )uw{+}2 \left (m{+}1\right )vw\Bigr] {i_2}\Biggr \},
\end{eqnarray}
\begin{eqnarray}
\label{betais2}
{\beta_v}&=&-(4-d)\Biggl\{v-\frac{1}{6}\left[8{v}^{2}{+} 2\left
(m+2\right )uv{+}2uw{+}2\left (m+1\right )vw{+}3{w}^{2 }\right] +
\frac{1}{9}\Bigl [44{v}^{3}+\Bigr.\nonumber\\ &&
 24\left
(m+2\right ) u{v}^{2}{+}2\left (3\,m{+}6\right
)v{u}^{2}{+}2\,\left (6\,m{+}24\right )uvw {+} 2\left
(12\,m+12\right )w{v}^{2}+ \nonumber\\ && \!\!\left (3m{+}45
\right )v{w}^{2}{+}2\left (m{+}8\right )u{w}^{2}\!+4{u}^{2}w{+}
\Bigl.\left (3m {+}9\right ){w}^{3}\Bigr ]{i_1}{+}
\frac{2}{9}\Bigl [ 2{v}^{3}{+}\left (m{+}2 \right
)v{u}^{2}{+}\Bigr.\nonumber\\ && 2\left (m{+}2\right
)u{v}^{2}{+}\Bigl. \Biggl. \frac{m{+}3}{2}{w}^{2}v{+}2\left (m{+}2
\right )uvw{+}2\left (m{+}1\right ){v}^{2}w\Bigr ]{i_2}\Biggr \},
\end{eqnarray}
\begin{eqnarray}
\label{betais3} {\beta_w}&=&\!-(4-d)\,w \Biggl
\{1-\frac{1}{6}\left[\left (m+4\right )w+12v+4u\right]
+\frac{1}{9}\Bigl [\left (5m+27\right ){w}^{2}{+}84{v}^{2}+
\Bigr.\Biggr. \nonumber\\ &&\!\Bigl. 4\left
(6m{+}15\right )vw{+}2\left
(5m{+}22\right )uw{+} 4\left (3m{+}18\right )uv{+} 2\left
(m{+}6\right ){u}^{2}\Bigr ] {i_1}{+} \frac{2}{9}\Bigl [2{v}^{2}{+}
\Bigr.\nonumber\\ && \!\Biggl.\Bigl.
 2\left (m{+}2\right )\!uv{+}
\left(m{+}2\right )\!{u}^{2}{+} \frac{ m{+}3}{2}{w}^{2}{+} 2\left
(m{+}2\right )\!uw{+}2\left (m{+}1\right )\!vw \Bigr ]{i_2}\!
\Biggr \}.
\end{eqnarray}
\begin{eqnarray}
\gamma_\phi&=&-\frac{(4-d)}{9}\left ({2}{v}^{2}+{2\left (m+2
\right )}uv+{\left (m+2\right
)}{u}^{2}+2\left(m+2\right)uw+\frac{m+3}{2}{w}^{2}+
\right.\nonumber\\ &&\Bigl.2\left(m+1\right)vw\Bigr){i_2},
\label{gammais1}
\\
\bar{\gamma}_{\phi^2}&=&\frac{(4-d)}{3}\Biggl
\{\frac{1}{2}\Bigl(2v+\left (m+2\right )u+\left(m+1\right)w\Bigr)-
(2{v}^{2}+2 \left (m+2\right )uv+\Biggr.\nonumber\\ &&
\Biggl.\left (m+2\right ){u}^{2}+
2\left(m+2\right)uw+\frac{m+3}{2}{w}^{2}+ 2\left(m+1\right)vw
){i_1}\Biggr\}, \label{gammais2}
\end{eqnarray}
Here, $u,v,w$ are the renormalized couplings, $i_1,i_2$ are the two-loop
integrals (c. f. formulas (\ref{2lmass1}), (\ref{2lmass2}) and comments
below them).

Again, in a first step of the analysis one may use the
$\varepsilon$-expansion. The first order $\varepsilon$-expansion
results were given in Ref. \cite{Aharony75}. They can be
reproduced from formulas (\ref{betais1}) -- (\ref{betais3}). In
particular one gets eight fixed points with the coordinates given
in Table~\ref{etabis} \cite{noteeps}.
The second order contributions in $\varepsilon$ to the fixed point
coordinates calculated in Ref.~\cite{Dudka01a} do not change
qualitatively the results obtained \cite{noteerr}. The main
question of interest now is whether the above described picture of
the runaway solution is not an artifact of an
$\varepsilon$-expansion. To shed light on this question one can
use a more refined analysis of the fixed points and their
stability, similar as it was done in the subsections \ref{VIA},
\ref{VIB}.

\begin{table}[htp]
\ttbl{30pc}{\label{etabis} Fixed points of the RAM with isotropic
distribution  of the random anisotropy axis in
$\varepsilon$-expansion. Here,
$x_\pm=(m-2\pm\protect\sqrt{(m-2)^2+48})/8$,
$y_\pm=(m-2-2mz
\pm$ $\protect\sqrt{(m-2-2mz)^2+4(12-8z)})/8$,
$z=(m+6)/(m+8)$. \newline}
{\begin{tabular}{|l|c|c|c|} \hline
  & $u^*$ & $v^*$ & $w^*$ \\ \hline
 I.&  0 & 0 & 0  \\ II.&    $\frac{6}{m{+}8}\varepsilon$&0 & 0\\
III. & 0&   $\frac{6}{8}\varepsilon$ & 0
\\ IV.
& $\frac{3}{2(m{-}1)}\varepsilon
$
&
$
-\frac{3(4{-}m)}{8(m{-}1)}\varepsilon $ & 0 \\ V. &
0&$\frac{6x_{+}}{12x_{+}+m+4}\varepsilon$
&$\frac{6}{12x_{+}+m+4}\varepsilon$
\\
VI. & 0&
$
\frac{6x_{-}}{12x_{-}+m+4}\varepsilon $  &
$
\frac{6}{12x_{-}+m+4}\varepsilon
$
\\
VII. & $-\frac{6z}{12y_{+}-4z+m+4}\varepsilon$
&$\frac{6y_{+}}{12y_{+}-4z+m+4}\varepsilon$&
$\frac{6}{12y_{+}-4z+m+4}\varepsilon$
\\
VIII. & $-\frac{6z}{12y_{-}-4z+m+4}\varepsilon$ &
$\frac{6y_{-}}{12y_{-}-4z+m+4}\varepsilon$ &
$\frac{6}{12y_{-}-4z+m+4}\varepsilon$
\\
   \hline
\end{tabular}}
\end{table}

\begin{figure}[htbp]
\begin{center}
{\input {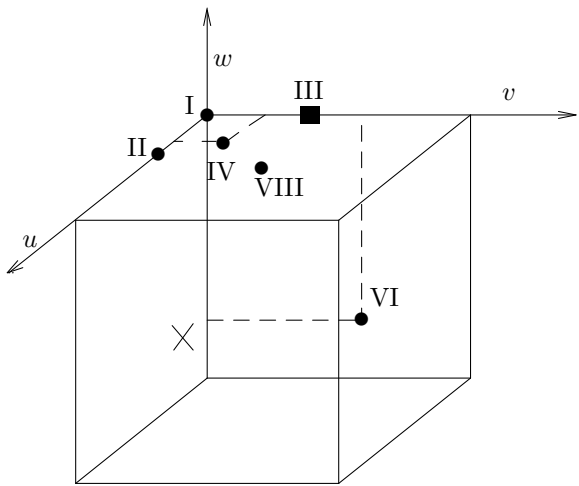}}
\end{center}
\vspace{-2.2cm} \caption{\label{ramfig1} Fixed points of the RAM
with isotropic distribution of a local anisotropy axis. The only
f\/ixed points located in the octant $u>0,v>0,w<0$ are shown.  The
filled box shows the stable fixed point, the cross denotes typical
initial values of the couplings.}
\end{figure}

All fixed points with $u>0,v>0,w<0$ appear to be unstable for
$\varepsilon>0$ except of the ``polymer" $O(n=0)$ fixed point III
which is stable for all $m$. However, the presence of a stable
fixed point is not a sufficient condition for the 2nd order phase
transition. The fixed point should be accessible from the initial
values of couplings and it is not the case for the location of
fixed points shown in Fig.~\ref{ramfig1}. Recall that the initial
conditions of the RAM effective Hamiltonian in our case are given
by $u_0>0$, $v_0>0$, $w_0<0$ and $w_0/v_0=-m$ (see subsection
\ref{IVC}).  Indeed, starting from the region of initial
conditions (denoted by a cross in the figure) for zero value of
$u$ one meets a separatrix joining the unstable fixed points I and
VI and the system cannot reach the stable fixed point III.  As far
as both fixed points I and VI are strongly unstable with respect
to $u$, the fixed point III is not accessible for arbitrary
positive $u$ either. Finally one concludes that the 2nd order
phase transition is absent in the model as shown by runaway
solutions of the RG equations.

Considering the two-loop RG functions
(\ref{betais1})--(\ref{betais3}) directly at $d=3$ and applying
the Pad\'e-Borel resummation (\ref{BorIm})--(\ref{res}) one gets
\cite{Dudka01a,Dudka} the fixed point coordinates given in the
Table \ref{ramtab1} \cite{noteerr}. Resummed two-loop results
qualitatively confirm the picture obtained in the first order in
$\varepsilon$-expansion: the stability of the fixed points does
not change after resummation. This supports a conjecture of
Aharony \cite{Aharony75} about the absence of an accessible stable
fixed point for the RAM with isotropic distribution of the local
anisotropy axis.

In the other fixed points one may recover the two-loop results for the
$O(m)$ model (fixed point II), the polymer $O(n=0)$ model (III) and
the diluted Stanley model (IV). The fixed
point VIII contains all three couplings but is both unstable and not
accessible from the initial values of couplings. The values of the
correlation length and pair correlation function critical exponents
$\nu$ and $\eta$, resummed in a similar way, are given in the table.
When calculated in unstable fixed points, they are considered as
effective exponents.

\begin{table}
\ttbl{30pc}{\label{ramtab1} Resummed values of the fixed points and
critical exponents for isotropic case in two-loop approximation
for $d=3$ \protect\cite{Dudka,noteerr}. \newline}
{\begin{tabular}{|c|c|c|c|c|c|c|}
   \hline
FP &$m$ & $u^*$ & $v^*$ & $w^*$ &
  $\nu$ &$\eta$\\
   \hline
I&$\forall m$ &0 &0 &0& &\\ \hline II&2 &0.9107&0  &0 &0.663
&0.027\\ &3 &0.8102 &0  &0 &0.693 &0.027\\ &4 &0.7275 &0  &0 &
0.724&0.027\\
 \hline
III&$\forall m$ &0 &1.1857 &0 &0.590 &0.023\\ \hline IV&2 &0.9454
&-0.0322 &0& 0.668&0.027\\ &3  &0.6460 &0.1733& 0&0.659 &0.027 \\
&4 &0.4851&0.2867&0&0.653 &0.028\\ \hline VI&2  &0 &1.4650
&-1.6278 &0.449 &-0.028\\
\hline
VIII&2& 0.7991 & 0.7311 & -0.5360 & 0.618 & 0.026 \\
&3& 0.8450 & 0.5934 & -0.4506 & 0.614 & 0.025\\
&4& 0.8762 & 0.5060 & -0.3956 & 0.612 & 0.025\\
\hline
\end{tabular}}
\end{table}

\subsubsection{A cubic distribution of the random anisotropy
axis}\label{VIC2}

Now we turn our attention to the  case of the RAM with a cubic
distribution of the random anisotropy axis (\ref{dist2}).
The RG functions that correspond to the effective Hamiltonian
(\ref{hamcub}) in two-loop approximation were obtained in
Ref.~\cite{Dudka01b} within the massive renormalization scheme
(\ref{conditions}). In the replica limit $n=0$ they read:
\begin{eqnarray}
{\beta_u}&=&-\varepsilon\,u\Biggl\{1-\frac{1}{6}\left[\left
(m{+}8\right )u{+}12v{+}4w{+}6y\right]{+}\frac{1}{9}\Bigl[
2\left(5m{+}22\right){u}^{2}{+}8\left(3m{+}15\right )vu{+}
 \Bigr.\Biggr.\nonumber\\
&&84{v}^{2}{+} 12{w}^{2}{+}68uw{+}72uy{+}18{y}^{2}{+} \Bigl.
72vw{+}108vy{+}36wy\Bigr] {i_1}{+} \frac{2}{9}\Bigl[\left
(m{+}2\right )\!{u}^{2}{+} \nonumber\\ &&\Biggl.\Bigl. 2\left
(m+2\right )uv{+}2{v}^{2}{+}
2{w}^{2}{+}6uw{+}4vw{+}6vy{+}6uy{+}6wy{+}3{y}^{2}\Bigr]{i_2}\Biggr
\}, \label{betacub1}
\end{eqnarray}
\begin{eqnarray}
{\beta_v}&=&-\varepsilon\Biggl\{v-\frac{1}{6}\left[8{v}^{2}{+}2\left
(m{+}2\right)uv{+}2uw{+}4vw{+}6vy\right]{+}\frac{1}{9}
\Bigl[44{v}^{3}{+}48{v}^{2}w{+} \Bigr.\Biggr. \nonumber\\
&&12{w}^{2}v{+}24\left (m+2 \right )u{v}^{2}{+}2\left
(3m{+}6\right )v{u}^{2}{+}
4{w}^{2}u{+}4{u}^{2}w{+}72{v}^{2}y{+}18{y}^{2}v{+} \nonumber\\
&&\Bigl.60uvw{+}36uvy{+}36vwy\Bigr] {i_1}+\frac{2}{9}\Bigl[
2{v}^{3}{+}2\left (m{+}2\right )u{v}^{2}{+}\left (m{+}2\right
)v{u}^{2}{+}\nonumber\\
&&\Biggl.\Bigl.2{w}^{2}v{+}6uvw{+}4{v}^{2}w{+}
6{v}^{2}y{+}6uvy{+}6vwy{+}3{y}^{2}v\Bigr] {i_2}\Biggr\},
\label{betacub2}
\end{eqnarray}
\begin{eqnarray}
{\beta_w}&=&-\varepsilon\,w\Biggl\{1-\frac{1}{6}\left[8w{+}12v{+}4u{+}
6y\right]+ \frac{1}{9}\Bigl[
44{w}^{2}{+}84{v}^{2}{+}120wv{+}2\left (m{+}6\right
){u}^{2}{+}\Bigr.\Biggr. \nonumber\\ &&\Bigl.
68uw{+}72wy{+}18{y}^{2}{+}2\left (6m+36\right )uv{+}108vy{+}36uy
\Bigr] {i_1}{+}\frac{2}{9}\Bigl[\left (m{+}2\right )\!{u}^{2}{+}
\nonumber\\ &&\Biggl.\Bigl. 2\left (m+2\right )uv{+}2{v}^{2}{+}
2{w}^{2}{+}6uw{+}4vw{+}6vy{+}6uy{+}6wy{+}3{y}^{2}\Bigr]{i_2}\Biggr
\}, \label{betacub3}
\end{eqnarray}
\begin{eqnarray}
{\beta_y}&=&-\varepsilon\Biggl\{ y-\frac{1}{6} \left[
9{y}^{2}{+}8uw{+}12vy{+}12uy{+}12wy\right
]{+}\frac{1}{9}\Bigl[\left (4m+72\right ){u}^{2}w{+}
72{w}^{2}u{+}\Bigr.\Biggr. \nonumber\\ &&
54{y}^{3}{+}84{v}^{2}y{+} \left(6m+84\right
){u}^{2}y{+}84{w}^{2}y{+}144{y}^{2}v{+}
144{y}^{2}u{+}144{y}^{2}w{+} \nonumber\\ && \Bigl.96uvw{+}2\left
(6m+84\right )uvy{+}252uwy{+}168vwy\Bigr]{i_1}
{+}\frac{2}{9}\Bigl[2{v}^{2}y{+}2\left (m+2\right )uvy{+}\Bigr.
\nonumber\\ && \Biggr.\Bigr. \left (m{+}2\right ){u}^{2}y{+}2
{w}^{2}y{+}6uwy{+}
4vwy{+}6{y}^{2}v{+}6{y}^{2}u{+}6{y}^{2}w{+}3{y}^{3}\Bigr]{i_2}\Biggr\}
, \label{betacub4} \\ \gamma_\phi&=&-\frac{\varepsilon}{9}\left
({2}{v}^{2}+{2\left (m+2 \right )}uv+{\left (m+2\right
)}{u}^{2}+6uw+2{w}^{2}+{4}vw+6vy+\right. \nonumber\\
&&\left.6uy+6wy+3{y}^{2}\right){i_2}, \label{gammacub1}
\\
\bar{\gamma}_{\phi^2}&=&\frac{\varepsilon}{3}\Biggl
\{\frac{1}{2}\Bigl(2v+\left (m+2\right )u+2w+y\Bigr)-\left
(2{v}^{2}+2 \left (m+2\right )uv+\left (m+2\right
){u}^{2}+\right.\Biggr.\nonumber\\ && \Biggl.\left.
6uw+2{w}^{2}+4vw+6vy+6uy+6wy+3{y}^{2} \right){i_1}\Biggr\},
\label{gammacub2}
\end{eqnarray}
Again, $i_1,\, i_2$ stand for the two-loop integrals and the
renormalized couplings $u$, $v$, $w$ and $y$ correspond to the effective
Hamiltonian (\ref{hamcub}).

As in the isotropic case the first RG result for the RAM with cubic
distribution of local anisotropy axis was obtained in the first order
in $\varepsilon$ \cite{Aharony75}. It brought about the evidence of 14
fixed points and may be easily reproduced from the functions
(\ref{betacub1})--(\ref{gammacub2}) removing the two-loop
contributions. The coordinates of the fixed points are listed in the
Table ~\ref{etabcub} \cite{noteeps}.

\begin{table}[!htbp]
\ttbl{30pc}{ \label{etabcub} Fixed points of the RAM
    with the random cubic anisotropy distribution
    ($\varepsilon$--expansion). Here,
    {$\alpha_\pm=(m-4\pm\protect\sqrt{m^2+48})/8$},
    {$\beta_\pm=-(m+12\pm\protect\sqrt{m^2+48})/6$},
    {$A_{\pm\pm}=6\alpha_\pm+3\beta_\pm+m+6$}. The fixed points
    XV-XVII appear only in the two-loop approximation due to the
    degeneracy of the corresponding one-loop functions.\newline}
{\begin{tabular}{|l|c|c|c|c|}
\hline & $u^*$ & $v^*$ & $w^*$ & $y^*$\\ \hline I.&  0 & 0 & 0 & 0
\\ II. & $\frac{6}{m+8}\varepsilon$&   0 & 0 & 0\\ III. & 0&
$\frac{6}{8}\varepsilon$ & 0 & 0 \\ IV.& 0 & 0 &
$\frac{6}{8}\varepsilon$ & 0
\\ V. & 0 & 0 & 0 &
$\frac{6}{9}\varepsilon$\\ VI.
 &
$\frac{6}{4(m-1)}\varepsilon$ &$\frac{6(m-4)}{16(m-1)}\varepsilon$
& 0 & 0 \\
 VII.&0 &$\frac{3}{2}\varepsilon$  &
$-\frac{3}{2}\varepsilon$& 0\\ VIII. & $\frac{2}{m}\varepsilon$&
0& 0 &$\frac{2(m-4)}{3m}\varepsilon$\\ IX. $m\neq 2$&
$\frac{1}{m-2}\varepsilon$& $\frac{m-4}{4(m-2)}\varepsilon$  & 0 &
$\frac{m-4}{3(m-2)}\varepsilon$\\ X. & 0& $\frac{1}{2}\varepsilon$
& $-\frac{1}{2}\varepsilon$ & $\frac{2}{3}\varepsilon$
\\
XI.& $\frac{3}{A_{++}}\varepsilon$&
$\frac{3\alpha_+}{A_{++}}\varepsilon$  &
$\frac{3(m+4)}{4A_{++}}\varepsilon$ &
$\frac{3\beta_+}{A_{++}}\varepsilon$\\
 XII.&
$\frac{3}{A_{+-}}\varepsilon$&
$\frac{3\alpha_+}{A_{+-}}\varepsilon$  &
$\frac{3(m+4)}{4A_{+-}}\varepsilon$ &
$\frac{3\beta_-}{A_{+-}}\varepsilon$\\ XIII.&
$\frac{3}{A_{-+}}\varepsilon$&
$\frac{3\alpha_-}{A_{-+}}\varepsilon$  &
$\frac{3(m+4)}{4A_{-+}}\varepsilon$ &
$\frac{3\beta_+}{A_{-+}}\varepsilon $\\ XIV.&
$\frac{3}{A_{--}}\varepsilon$&
$\frac{3\alpha_-}{A_{--}}\varepsilon$  &
$\frac{3(m+4)}{4A_{--}}\varepsilon$ &
$\frac{3\beta_-}{A_{--}}\varepsilon$
\\
\hline XV.& 0 & 0 & $\mp\sqrt{\frac{54}{53}\varepsilon}$ &
$\pm\frac{4}{3}\sqrt{\frac{54}{53}\varepsilon}$ \\ XVI.
&0&$\mp\sqrt{\frac{54}{53}\varepsilon}$&  0 &
$\pm\frac{4}{3}\sqrt{\frac{54}{53}\varepsilon}$
\\
XVII.{\scriptsize $m=2$}  & $\mp2\sqrt{\frac{54}{53}\varepsilon}$
& $\pm\sqrt{\frac{54}{53}\varepsilon}$& 0 &
$\pm\frac{4}{3}\sqrt{\frac{54}{53}\varepsilon}$ \\ \hline
\end{tabular}}
\end{table}

\begin{figure} [htbp]
\begin{center}
{\input {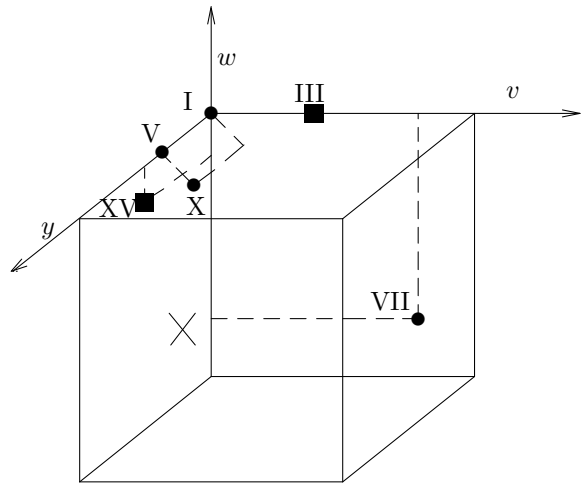}}
\end{center}
\vspace{-2.0cm} \caption{ \label{ramfig2} Fixed points of the RAM
with distribution of a local anisotropy axis along hypercube axis
for $u=0$. The only fixed points located in the region $v>0,w<0$
are shown. Filled boxes show the stable fixed points, the cross
denotes typical initial values of couplings.}
\end{figure}

The results of the linear $\varepsilon$ analysis \cite{Aharony75}
states that among the fixed points with $u>0,v>0,w<0$ only a
``polymer" $O(n=0)$ fixed point III is stable for all $m$ for
$\varepsilon>0$, but it is not reachable from the initial values
of couplings (see Fig. \ref{ramfig2}). The reason is a separatrix
joining the unstable fixed points I and VII and separating initial
values of couplings (shown by a cross in the Fig. \ref{ramfig2})
and fixed point III. The possible runaway behavior of the RG flow
lead to the conclusion about a smearing out of the phase
transition as $T_c$ is approached \cite{Aharony75}.
However, a subsequent study of Mukamel and Grinstein
\cite{Mukamel82} brought about the possibility of a second order
phase transition with the scenario of a weakly diluted quenched
Ising model \cite{Grinstein76}. Indeed, accounting for the
$\varepsilon^2$ terms qualitatively changes the picture:
performing the perturbation theory expansion one gets
\cite{Dudka01b,noteerr} not only the corrections to the
coordinates of the fixed points I--XIV (we do not display them
here as far as they as are too cumbersome) but also the new fixed
points XV, XVI, XVII (see the bottom of the Table \ref{etabcub}).
The appearance of the pairs of fixed points denoted by XV and XVI
is caused by the fact that the $\beta$-functions $\beta_w$,
$\beta_y$ at $u=v=0$ ($\beta_u$, $\beta_v$ at $w=y=0$,
correspondingly) are degenerated at the one loop level. This leads
to the $\sqrt\varepsilon$-expansion (c.f.  subsection~\ref{VIA})
of the weakly diluted quenched Ising model. The $\sqrt\varepsilon$
expansion of the fixed point XVII holds only for $m=2$ and is
caused by the one-loop degeneracy of the $\beta_u$, $\beta_v$,
$\beta_y$ functions for $w=0$ (c. f. singularity at $m=2$ in the
$\varepsilon$-expansion of the fixed point IX)~\cite{Dudka01b}.

Checking the stability of the new fixed points one finds that all of
them are unstable except the Ising-like fixed point with
$w<0,\,y>0$ of the pair XV. Moreover, this point is reachable
from the initial values of the couplings. As far as this is the
fixed point of the diluted Ising model one concludes, that in the
critical region the RAM with cubic distribution of the random anisotropy
axis (\ref{dist2}) decouples into $m$ independent dilute Ising
models and the phase transition for any $m$ is governed by the
familiar random Ising model critical exponents. However, it is
worthwhile to keep in mind that above picture is obtained in the
frames of the ``naive" analysis of the $\varepsilon$ (and
$\sqrt\varepsilon$) expansion and it is highly desirable to
confirm it by a more reliable analysis of fixed points and their
stability.

The results of the application of the Pad\'e-Borel resummation
procedure (\ref{BorIm})--(\ref{res}) to the two-loop $\beta$-functions
(\ref{betacub1})--(\ref{betacub4}) at $d=3$ are given in Refs.
\cite{Dudka,Dudka01b} where 16 fixed points were obtained for this
model. In the Table~\ref{ramtab2} we present the numerical values of
fixed points coordinates with $u^*>0$, $v^*>0$, $w^*<0$.  The last
fixed point XV in Table~\ref{ramtab2} corresponds to the stable fixed
point of $\sqrt\varepsilon$-expansion. It has coordinates with
$u^*=v^*=0$, $w^*<0$ and $y^*>0$ and is accessible from the typical
initial values of couplings (marked by a cross in the Fig.
\ref{ramfig2}). The Pad\'e-Borel resummation procedure
(\ref{betacub1})--(\ref{betacub4}) does not reveal any other stable
fixed points in the region of interest. Thus one draws the conclusion
that the system in the critical regime is characterized by the same
set of critical exponents as the weakly diluted  quenched
Ising model.

In the other fixed points, one recovers the familiar two-loop
numerical results for the Gaussian (fixed points I, VII),
$m$-vector (II), polymer \mbox{$O(n=0)$} (III),
Ising (V, X), diluted $m$-vector (VI),
and cubic (VIII) models. Fixed point IX belongs to a
new universality class. In the Table \ref{ramtab2} we give the
numerical values of the critical exponents in these fixed points
as well: if the flow from the initial values of couplings passes
near these fixed points one may observe an effective critical
behavior governed by these critical exponents. Note that the
fixed point VI has a singularity for $m=1$ which formally leads to
the $\sqrt \varepsilon$ expansion. However, as it follows
already from the microscopic Hamiltonian (\ref{ram}) for $m=1$ the RAM
degenerates to the regular Ising model. This also may be seen from
the effective Hamiltonians (\ref{hamis}), (\ref{hamcub}): in the
replica limit at $m=1$  $v$ and $w$ terms (and for (\ref{hamcub})
also $u$ and $y$ terms) appear to be of the same symmetry.
Furthermore, couplings $v$ and $w$ appear to have opposite signs and
the same absolute value, therefore they are canceled and the problem
reduces to the scalar $\phi^4$ theory with only one coupling
\cite{degen}.

As a possible generalization of the RAM one may consider the case
when the quen\-ched randomness is present in both the random-site and
the random-anisotropy form. This is described by effective
Hamiltonians (\ref{hamis}), (\ref{hamcub}) where the coupling
$v_0$ may be of either sign. We have checked the region $v<0$ for
the presence of new fixed points and verified that they are
absent. Therefore, again the fixed point XV is the only reachable and
stable one and the observed critical behavior is unique. As an
interesting extension of these studies one may consider the
influence of other types of disorder in combination with the
random anisotropy as well as of other random axis distributions.

\begin{table}[htbp]
\ttbl{30pc}
{\label{ramtab2} Resummed values of the fixed points and critical
exponents for cubic distribution
in two-loop approximation for $d=3$ \protect\cite{Dudka,Dudka01b}.
\newline
}
   {\begin{tabular}{|c|c|c|c|c|c|c|c|}
   \hline
FP &$m$ & $u^*$ & $v^*$ & $w^*$ & $y^*$ &  $\nu$ &$\eta$\\
   \hline
   I&$\forall m$ &0 &0 &0&0 &1/2 &0 \\
   \hline
 &2 &0.9107 &0  &0& 0 &0.663 &0.027\\
II&3 &0.8102&0  &0 & 0&0.693 &0.027\\ &4 &0.7275 &0  &0 &
0&0.720&0.026\\ \hline
 III&$\forall m$  &0 &1.1857 &0& 0 &0.590 &0.023 \\
 \hline
V&$\forall m$ &0 &0 &0 &1.0339&0.628 &0.026\\ \hline
 VI&3
&0.6460 &0.1733 &0 & 0&  0.659&0.027\\ &4 &0.4851&0.2867  & 0&
0&0.653 &0.027\\ \hline
 VII&$\forall m$  &0&2.1112 &-2.1112 &0&1/2 &0\\ \hline
 &2 &1.5508 &0  &0 &-1.0339& 0.628 &0.026\\
 VIII&3  &0.8393&0 &0 &-0.0485& 0.693 &0.027\\
 &4  &0.5259&0 &0 &0.3624& 0.709 &0.026\\
\hline
 IX&3 &0.7096 &0.1695 & 0& -0.1022&0.659& 0.027\\
 &4  &0.4190&0.2751 & 0& 0.1432&0.653 &0.027\\
\hline X&$\forall m$  &0&0.6678 &-0.6678 &1.0339&0.628 &0.026\\
\hline XV&$\forall m$ &0&0&-0.4401&1.5933&0.676&0.031\\ \hline
\end{tabular}}
\end{table}

As we have seen from the analysis of the subsection \ref{VIC} the
local anisotropy axis distribution has a crucial influence on the
scenarios of low-temperature behavior. For the RAM with isotropic
distribution the inaccessibility of any stable fixed point leads to
the conclusion that a second order phase transition is absent. The
existence of a reachable disordered Ising-like fixed point in case of
the RAM with cubic distribution of the random anisotropy axis
indicates that the system undergoes a second order phase transition
with the critical exponents of the weakly  diluted quenched Ising
model. Here, we want to attract the readers attention to the
similarity of the critical behavior of both the
random-site \cite{Grinstein76} and the random-anisotropy \cite{Harris73}
quenched magnets: if new critical behavior is present then it
is always governed by the critical exponents of the site-diluted
Ising type. The above calculations of a ``phase diagram" of the RAM
are based on two-loop expansions improved by a resummation technique.
Once the qualitative picture becomes clear there is no need to go into
higher orders of a perturbation theory as far as the critical
exponents of site-diluted Ising model are known by now with high
accuracy (see Table ~\ref{table_theory} and discussion of the subsection
~\ref{VIA}).

\section{Conclusions}\label{VII}

To conclude this review, let us return to the question posed in
the introduction: what information about the influence of weak quenched
disorder on criticality can one get analyzing the RG expansions?
Comparing the results of the theoretical analysis in Section \ref{VI} with
the available data about different physical phenomena listed in
Section \ref{III} one definitely can state that in most cases the RG
analysis provides a correct description. Indeed, the change of the
universality class of the Ising model upon
weak quenched dilution predicted by the RG analysis is confirmed by
numerous experiments on diluted uniaxial magnets ${\rm
  Fe_pZn_{1-p}F_2}$, ${\rm Mn_pZn_{1-p}F_2}$ as well as by extensive
MC simulations (c. f. data of the Table \ref{table_theory} and of
the Section \ref{III}). By high precision experiments under
microgravity conditions it was shown \cite{He4dil} that the
specific heat asymptotic critical exponent of liquid ${\rm He^4}$
for the $\lambda$-transition is not influenced by quenched
disorder and belongs to the universality class of the $m=2$
Stanley model. This result also follows from the RG analysis as
far as the Stanley model heat capacity critical exponent $\alpha$
changes its sign already for $m<2$ \cite{alpha}. The MC
simulations of self-avoiding walks on random-site lattices
\cite{Barat95,ransaws} confirmed the theoretical result of Ref.
\cite{Kim} that universal properties of self-avoiding walks are
not influenced by point-like uncorrelated disorder. The change of
the universality class of self-avoiding walks by a
long-range-correlated disorder predicted by the RG analysis
\cite{Blavats'ka01a,Blavats'ka01} waits for an experimental or
simulational confirmation. The MC study of the influence of the
long-range-correlated disorder on the ferromagnetic phase
transition confirms the change of the universality class in this
case \cite{Ballesteros99}. One more interesting check of the RG
predictions might be the verification (experimental or in MC
simulations) of the envisaged random Ising model critical behavior
for the random anisotropy magnet with a cubic distribution of the
anisotropy axis \cite{Dudka,Dudka01b,Mukamel82,Korzhenevskii88}.

As we have seen in these lectures, for the majority of the
phenomena listed above a careful analysis of the RG expansions
provides also a quantitative description, giving reliable numbers
for the physical values describing criticality. However, if one
compares the RG expansions for non-diluted (``pure") and diluted
systems one certainly notices essential differences. A typical
scenario that one encounters when analyzing the critical
properties of the $d=3$ Stanley model (i.e. applying the field
theoretical RG technique to the $O(m)$-symmetrical $\phi^4$ model)
may be sketched by two steps: (i) already in the first non-trivial
order of the perturbation theory one observes a new non-trivial
critical behavior (a stable non-trivial fixed point); (ii) the
higher order contributions do not change the picture qualitatively
and improve the numerical outcome provided the resummation is
applied. This scenario does not hold for most of the models with
disorder. Here, the first-order perturbation theory expansions as
a rule do not lead to new criticality (a non-trivial fixed point
is absent or its stability is questioned). The (experimentally
confirmed!) answer about new critical behavior is obtained only by
considering higher order contributions with the help of the
resummation technique. But then again, even the resummed series
appear to be divergent starting from some order of perturbation
theory showing an ``optimal truncation" behavior. Some expansions
that are familiar and useful for the ``pure" models appear to be
of no use for certain disordered models: e.g. the
$\varepsilon$-expansion turns into the $\sqrt
\varepsilon$-expansion for the random-site Ising model and does
not lead to any quantitative information.

Technically, this ``strange" behavior of the perturbation theory
expansions for the disordered models is not strange: it rather signals
the divergent nature of the series. For some disordered models this
behavior was shown \cite{nonsum,Alvarez00} to originate from
Griffith-like singularities \cite{Griffiths69} caused by zeroes of
the partition function of the pure system. The physical reason for
Griffith singularities is the existence of ordered islands within the
temperature interval below the critical temperature of the pure model
and above that of the diluted model.  So in this sense the studies of
random models reviewed above may serve as examples when more
complicated physics leads to a more complicated theoretical
description. And as one may conclude from the results obtained so far
in many cases this description is a fruitful one.

\section*{Acknowledgments}

The materials presented here were the subject of lectures given by one
of us in summer semester 2001 in the Universit\'e Henri Poincar\'e,
Nancy 1 (France) and at the 2nd Pamporovo workshop on cooperative
phenomena in condensed matter (August 2001, Bulgaria).
Yu.H. expresses his thanks to Bertrand Berche and
colleagues at the Laboratoire de physique des mat\'eriaux for
their hospitality, support and many useful comments. It is also
a special pleasure to express our gratitude to Dimo I.
Uzunov and his colleagues from the G. Nadjakov Institute of Solid
State Physics  of  the Bulgarian Acad. Sci. for their devoted work
on organization of Pamporovo workshops.

M. D. acknowledges the Ernst Mach research fellowship of the
\"Osterreichisher Austauschdienst. This work
was supported in part by \"Osterreichische Nationalbank
Jubil\-\"aums\-fonds through grant No 7694.


\begin{thebibliography}{150}

\bibitem{Forgacs91}
G. Forgacs, R. Lipowsky, Th. Nieuwenhuizen, in
{\it Phase Transitions and Critical Phenomena},
ed. by C.~Domb and J.~L.~Lebowitz, ( Academic Press, London,
1991), vol. 14.

\bibitem{rgbooks}
E. Br\'ezin, J. C. Le Guillou, J. Zinn-Justin,
in {\it Phase Transitions and Critical Phenomena},
ed. by C.~Domb and M.~S.~Green, ( Academic Press, London, 1976),
vol. 6; D.~J.~Amit,  {\it Field Theory, the Renormalization Group, and
Critical Phenomena}, (World Scientific, Singapore, 1989);
J.~Zinn-Justin, {\it Quantum Field Theory and Critical
Phenomena}, (Oxford University Press, 1996);
H. Kleinert, V. Schulte-Frohlinde,  {\it Critical Properties of
$\phi^4$-Theories}, (World Scientific, Singapore, 2001).


\bibitem{Stanley68}
H.~E.~Stanley,
{\it Phys. Rev. Lett} {\bf 20}, 589 (1968)

\bibitem{d2}
Absence of the phase transition with spontaneous order parameter
for the systems with continuous symmetry and a short range
interaction at $d=2$ follows from the Mermin-Wagner theorem:
N.~D.~Mermin and N.~Wagner, {\it Phys. Rev. Lett.} {\bf 17}, 1133
(1966); exact solutions for 2d Stanley model for $m=4$ and $m=3$ also
demonstrate the absence of magnetic ordering: A.~M.~Polyakov and
P.~B.~Wiegmann, {\it Phys. Lett.} {\bf B131}, 121 (1983);
P.~B.~Wiegmann, {\it Pis'ma Zh. Eksp. Teor. Fiz.} {\bf 41}, 79 (1985)
[JETP Letters {\bf 41}, 95 (1985)].

\bibitem{Stanley} See e.g. H. E. Stanley, {\it Introduction to Phase
Transitions and Critical Phenomena}, (Claredon Press, Oxford
1971).

\bibitem{Guida98}
R. Guida, J. Zinn-Justin {\it J. Phys. A} {\bf 31}, 8103 (1998).

\bibitem{note2}
One can account the presence of non--magnetic impurities introducing
random interaction $J$. However, the corresponding random--bond model
is conjectured to exhibit universal critical behavior identical to the
site--diluted model: K.~Hukushima {\it J. Phys. Soc. Jap.} {\bf
69}, 631 (2000); P.~E.~Berche, C.~Chatelain, B.~Berche, W.~Janke, in
{\it MECO25, Middle European Cooperation in Statistical Physics},
(Pont-\`a-Mousson, France, 2000), p.51.

\bibitem{perc} See e.g. J. W. Essam,
in {\it Phase Transitions and Critical Phenomena},
ed. by C.~Domb and M.~S.~Green,  (Academic Press, London, 1972),
vol. 2.

\bibitem{Brout59}
R. Brout,
{\it Phys. Rev.} {\bf 115}, 824 (1959).

\bibitem{Fisher68}
M.~E.~Fisher,
{\it Phys. Rev.} {\bf 176}, 257 (1968).

\bibitem{Harris74}
A.~B.~Harris
{\it J. Phys. C} {\bf 7}, 1671 (1974).

\bibitem{note}
One should be careful in applying the Harris criterion ``naively"
to the SAW problem: see discussion in the subsection \ref{IIIE}.

\bibitem{Chayes86}
J. T. Chayes, L. Chayes, D. S. Fisher, T. Spenser
{\it Phys. Rev. Lett.} {\bf 57}, 2999 (1986).

\bibitem{Weinrib}
A. Weinrib and  B.I.  Halperin,
{\it Phys. Rev.} {\bf B27}, 413 (1983).

\bibitem{Dorogovtsev80}  S. M. Dorogovtsev,  {\it Fiz.
Tverd. Tela} (Leningrad) 22:321 (1980) [Sov. Phys.--Solid
State22:188 (1980)]

\bibitem {Korzhenevskii96}
A. L. Korzhenevsky, K. Herrmanns, W. Schirmacher,
{\it Phys. Rev. B53:14834} (1996).

\bibitem {Yamazaki88}
Y. Yamazaki, A. Holz, M. Ochiai, and Y. Fukuda, {\it Physica A}
{\bf 150}, 576 (1988).

\bibitem{Harris73} R. Harris, M.
Plischke, and M. J. Zuckermann, {\it Phys. Rev. Lett.} {\bf 31}, 160
(1973).

\bibitem{Aharony75}
A. Aharony,
{\it Phys. Rev.} {\bf B12}, 1038 (1975).

\bibitem{Imry75}
Y. Imry, S.-k. Ma, {\it Phys. Rev. Lett.} {\bf 35}, 1399 (1975).

\bibitem{Jayaprakash80}
See e. g. C. Jayaprakash and S.
Kirkpatrick, {\it Phys.
Rev.} {\bf B21}, 4072 (1980) and Y. Y. Goldschmidt
and A.  Aharony, {\it Phys. Rev.} {\bf B32}, 264 (1985).

\bibitem{Pelcovits78}
R. A. Pelcovits, E. Pytte, and
J. Rudnick, {\it Phys. Rev. Lett.} {\bf 40}, 476 (1978).


\bibitem{Dunlap81}
R.~A.~Dunlap and A.~M.~Gottlieb, {\it Phys. Rev.} {\bf B23}, 6106
(1981).

\bibitem{Birgeneau83}
R.~J.~Birgeneau, R.~A.~Cowley, G.~Shirane, H.~Joshizawa,
D.~P.~Belanger, A.~R.~King, and V.~Jaccarino,
{\it Phys. Rev.} {\bf B27}, 6747 (1983).

\bibitem{Belanger86}
D.~P.~Belanger, A.~R.~King, and V.~Jaccarino,
Phys. Rev. B {\bf 34}, 452 (1986).

\bibitem{Mitchell86}
P.~W.~Mitchell, R.~A.~Cowley, H.~Yoshizawa, P.~B\"oni, Y.~J.~Uemura,
and R.~J.~Birgeneau, {\it Phys. Rev.} {\bf B34}, 4719 (1986).

\bibitem{Barret86}
P.~H.~Barrett,
{\it Phys. Rev.} {\bf B34}, 3513 (1986).

\bibitem{Thurston88}
T.~R.~Thurston, C.~J.~Peters, R.~J.~Birgeneau, and P.~M.~Horn,
{\it Phys. Rev.} {\bf B37}, 9559 (1988).

\bibitem{Rosov88}
N.~Rosov, A.~Kleinhammes, P.~Lidbj\"ork, C.~Hohenemser, and
M.~Eibsch\"utz, {\it Phys. Rev.} {\bf B37}, 3265 (1988).

\bibitem{ranfield}
S.~Fishman and A.~Aharony, {\it J. Phys. C} {\bf 12}, L729 (1979);
J.~Cardy, {\it Phys. Rev.} {\bf B29}, 505 (1984).

\bibitem{withH}
C.~A.~Ramos, A.~R.~King, and V.~Jaccarino,
{\it Phys. Rev.} {\bf B37}, 5483 (1988);
I.~B.~Ferreira, A.~R.~King, and V.~Jaccarino,
{\it Phys. Rev.} {\bf B43}, 10797 (1991);
D.~P.~Belanger, J.~Wang, Z.~Slanic, S.-J.~Han, R.~M.~Nicklow, M.~Lui,
C.~A.~Ramos, and D.~Lederman,
{\it Journ. Magn. Magn. Mater.} {\bf 140}, 1549 (1995),
{\it Phys. Rev.} {\bf B54}, 3420 (1996);
J.~P.~Hill, Q.~Feng, Q.~J.~Harris, R.~J.~Birgeneau, A.~P.~Ramirez,
and A.~Cassanho,
{\it Phys. Rev.} {\bf B55}, 356 (1997);
Z.~Slani\v{c}, D.~P.~Belanger, and J.~A.~Fernandez-Baca,
{\it J. Magn. Magn. Mater.} {\bf 177}, 171 (1998);
Z.~Slani\v{c} and D.~P.~Belanger,
{\it J. Magn. Magn. Mater.} {\bf 186}, 65 (1998),
{\it Phys. Rev. Lett.}  {\bf 82}, 426 (1999).

\bibitem{Folk01} R. Folk, Yu. Holovatch, T. Yavors'kii,
{\it cond-mat/0106468} (2001), to appear in {\it Physics Uspiekhi} (2002).

\bibitem{ammagn}
H. Yamamoto, H. Onodera, K. Hosoyama, T. Masumoto, and H. Yamauchi,
{\it J. Magn. Magn. Mater.} {\bf 31-34}, 1579 (1983);
K. Winschuh and M. Rosenberg, {\it J. Appl. Phys.} {\bf 61}, 4401 (1987);
G. K. Nicolaides, G. C. Hadjipanayis, and K. V. Rao, {\it Phys. Rev. B}
{\bf 48}, 12759 (1993);
P.~D.~Babu and S. N. Kaul, {\it J. Phys.: Cond. Matt.}  {\bf 9}, 7189
(1997).

\bibitem{Perumal01}
A. Perumal, V. Srinivas, K. S. Kim, S. C. Yu, V. V. Rao, and R. A. Dunlap,
{\it J. Magn. Magn. Mater.}  {\bf 233}, 280 (2001).

\bibitem{magglass}
M. F\"ahnle, G Herzer, H. Kronm\"uller, R. Meyer, M. Saile, and T.
Egami, {\it J. Magn. Magn. Mater.}  {\bf 38}, 240 (1983);
S.~N.~Kaul, {\it Phys. Rev. B}   {\bf 38}, 9178 (1988);
S.~N.~Kaul and Ch. V. Mohan, {\it Phys. Rev. B}   {\bf 50}, 6157 (1994).

\bibitem{Kellner86}
W.-U. Kellner, T. Albrecht, M. F\"ahnle, and H. Kronm\"uller,
{\it J. Magn. Magn. Mater.} {\bf 62}, 169 (1986).

\bibitem{Boxberg94}
O.~Boxberg and K. Westerholt, {\it Phys. Rev. B}   {\bf 50}, 9331 (1994).

\bibitem{Westerholt}
K. Westerholt and G. Sobotta, {\it J. Phys. F}  {\bf 13}, 2371 (1983);
K. Westerholt, H. Bach, and R. R\"omer, {\it J. Magn. Magn. Mater.}  {\bf
45}, 252 (1984);
K. Westerholt, {\it Physica}  {\bf 130B}, 533 (1985);
K. Westerholt, {\it J. Magn. Magn. Mater.}  {\bf 66}, 253 (1987).

\bibitem{Fahnle}
M. F\"ahnle,
{\it J. Magn. Magn. Mater.}, {\bf  45}, 279 (1984).

\bibitem{Lviv}
M. Dudka, R. Folk, Yu. Holovatch, and D. Ivaneiko,
{\it J. Magn. Magn. Mater.}, (2002), to appear.

\bibitem{note4}
Only in a special case liquids in porous media are an example for
the site diluted Ising model; otherwise they are conjectured to be
examples for the random-field models
[ see e.g. E. Pitard, M. L. Rosinberg, G. Stell, G. Tarjus,
{\it Phys. Rev. Lett.} {\bf 74}, 4361 (1995)]

\bibitem{He4dil} J. Yoon, M. H. W. Chan, {\it Phys. Rev. Lett.}
{\bf 78}, 4801 (1997);
see also: G.~M.~Zassenhaus, J.~D.~Reppy   {\it Phys. Rev. Lett.}
{\bf 83}, 4800 (1999).

\bibitem{note3} Experimentally measured value of the heat capacity
critical exponent at the $\lambda$-transition in helium-4 reads:
$\alpha=-0.01056 \pm 0.00038$
[J.~A.~Lipa, D.~R.~Swanson, J.~A.~Nissen, Z.~K.~Geng, P.~R.~Williamson,
D.~A.~Stricker, T.~C.~P.~Chui, U.~E.~Israelsson, M.~Larson,
{\it Phys. Rev. Lett.}  {\bf 84}, 4894
(1999)]. Within the error bars $\alpha<0$ therefore due to the
Harris criterion a weak quenched disorder should not influence
the critical exponents.

\bibitem{Pelissetto00}
A. Pelissetto and E. Vicari, {\it cond-mat/0012164} (2000), to appear
in {\it Physics Reports} (2002).

\bibitem{Landau80}
D.~P.~Landau,
{\it Phys. Rev. B} {\bf 22}, 2450 (1980).

\bibitem{Marro86}
J.~Marro, A.~Labarta, and J.~Tejada,
{\it Phys. Rev.} {\bf B34}, 347 (1986); D.~Chowdhury and D.~Staufer,
{\it J. Stat. Phys.}  {\bf 44}, 203 (1986).

\bibitem{Wang89}
J.-S.~Wang and D.~Chowdhury,
{\it J. Phys. France} {\bf 50}, 2905 (1989).

\bibitem{Wang90}
J.-S.~Wang, M.~W\"ohlert, H.~M\"uhlenbein, and D.~Chowdhury,
{\it Physica A}  {\bf 166}, 173 (1990).

\bibitem{Heuer90}
H.-O.~Heuer,
{\it Europhys. Lett.} {\bf 12}, 551 (1990);
H.-O.~Heuer,
{\it Phys. Rev.} {\bf B42}, 6476 (1990).

\bibitem{Heuer93}
H.-O.~Heuer,
{\it J. Phys. A}  {\bf 26}, L333 (1993).

\bibitem{Parisi98}
H.~G.~Ballesteros, L.~A.~Fern\'andez, V.~Mart\'in-Mayor,
A.~Mu\~noz~Sudupe, G.~Parisi, and J.~J.~Ruiz-Lorenzo,
{\it Phys. Rev.} {\bf B58}, 2740 (1998).

\bibitem{Folk00}
R. Folk, Yu. Holovatch and
T. Yavorskii, {\it Phys. Rev.} {\bf B61}, 15114 (2000).

\bibitem {Yamazaki}
 Y. Yamazaki, A. Holz, M. Ochiai, Y. Fukuda, {\it Phys. Rev.}
{\bf B33}, 3460 (1985);
 Y.~Yamazaki, Y.~Fukuda, A.~Holz, M.~Ochiai,  {\it Physica A}
{\bf 136}, 303 (1986).

\bibitem{Cardy}
D. Boyanovsky and J. L. Cardy, {\it Phys. Rev.} {\bf B26}, 154 (1982).

\bibitem{Ballesteros99}
H. G. Ballesteros and G. Parisi, {\it Phys. Rev.} {\bf B60}, 12912
(1999).

\bibitem{Marques00}
M. Marqu\'es, J. Gonzalo and J. \'Iniquez, {\it Phys. Rev.}
{\bf E62}, 191 (2000).

\bibitem{Lee92}
J. C. Lee, R. Gibbs,
{\it Phys. Rev. B}45:2217 (1992).


\bibitem{desCloizeaux90}
J. des Cloizeaux, G. Jannink, {\it Polymers in Solution}, (Clarendon
Press, Oxford,  1990); L. Sch\"afer, {\it Excluded Volume Effects in
Polymer Solutions} (Springer, Berlin, 1999).

\bibitem{deGennes79}
P.-G. de~Gennes, {\it Scaling Concepts in Polymer
Physics}, (Cornell University Press, Ithaca and London, 1979).

\bibitem{Chakrabarti}
B. K. Chakrabarti and  J. Kert\'{e}sz,
{\it Z. Phys. B:Condensed Matter} {\bf 44}, 221 (1981).

\bibitem{Harris83}
 A. B. Harris,  {\it Z. Phys. B} {\bf 49}, 347 (1983).

\bibitem{Kim}
Y. Kim, {\it  J. Phys. C} {\bf 16}, 1345 (1983).

\bibitem{Meir89}
Y. Meir and A. B. Harris, {\it Phys. Rev. Lett.}  {\bf 63}, 2819
(1989).

\bibitem{Grassberger93} P. Grassberger, {\it J. Phys. A}
{\bf 26}, 1023 (1993).

\bibitem{Barat95}
K. Barat and  B. K. Chakrabarti, {\it Phys. Reports}  {\bf 258}, 377
(1995).

\bibitem{randomsaws}
S. Lee and H. Nakanishi, {\it Phys. Rev. Lett.}  {\bf 61}, 2022
(1987); S. Lee,  H. Nakanishi and Y.  Kim,  {\it Phys. Rev.} {\bf
B39}, 9561 (1988); J. Machta and T.R. Kirkpatrick,  {\it  Phys. Rev.}
{\bf A41}, 5345  (1990); A.V. Izyumov and K. B. Samokhin, {\it J.
Phys. A} {\bf 32}, 7843 (1999); A. R. Altenberger,  I. I. Siepmann,
 and  J. S. Dahler, {\it  J. Phys. A} {\bf 272}, 22 (1999); Y.
Shiferaw and Y. Y. Goldschmidt, {\it J. Phys. A} 33:4461 (2000).

\bibitem{Ordemann} For recent references see e.g. A. Ordemann,
M.~Porto, H.~Eduardo Roman, S.~Havlin, and A.~Bunde, {\it Phys.
Rev.} {\bf E61}, 6858 (2000); A.~Ordemann, M.~Porto, H.~Eduardo Roman,
and S.~Havlin, {\it Phys. Rev.} {\bf E63}, 020104(R) (2001).

\bibitem{ransaws}
K. Kremer, {\it Z. Phys. B}  {\bf 45}, 149 (1981);
S. Lee, and N. Nakanishi, {\it Phys. Rev. Lett.}  {\bf 61}, 2022
(1988).

\bibitem{Blavats'ka01a}
V. Blavats'ka, C. von Ferber, and Yu. Holovatch, {\it Phys. Rev}
{\bf E64}, 041102 (2001).

\bibitem{Blavats'ka01}
V. Blavats'ka, C. von Ferber, and Yu. Holovatch, {\it J. Mol. Liq.}
{\bf 91}, 77  (2001).

\bibitem{Cochrane78}
R. W. Cochrane, R. Harris, and M. J. Zuckermann, {\it Phys. Rep.}
{\bf 48}, 1 (1978).

\bibitem{Cochrane74}
R. W. Cochrane, R. Harris, and M. Plischke, {\it J. Non.-Cryst. Solids}
{\bf 15}, 239 (1974).

\bibitem{Cochrane75}
R. W. Cochrane, R. Harris, M. Plischke, D. Zobin, and M. J. Zuckermann,
{\it J. Phys. F} {\bf 5}, 763 (1975).

\bibitem{Chakrabati98}
J. Chakrabati, {\it Phys. Rev. Lett}  {\bf 81}, 385 (1998).

\bibitem{Wu92}
X.-l. Wu, W. I. Goldburg, M. X. Liu, and J.~Z.~Xue {\it Phys. Rev. Lett.}
{\bf 69}, 470 (1992).


\bibitem{Grinstein76}
G. Grinstein, A. Luther,
{\it Phys. Rev.} {\bf B13}, 1329 (1976)

\bibitem{replicas}
V. J. Emery,  {\it Phys. Rev.} {\bf B11}, 239 (1975); S. F. Edwards,
P.  W. Anderson {\it J. Phys. F} {\bf 5}, 965 (1975).

\bibitem{Dimo} One more extension of this model of a random
magnet is to take into account both influence of a single-ion
anisotropy  and random impurities as is done e.g. in I. D. Lawrie,
Y. T. Millev, and D. I. Uzunov {\it J. Phys. A} {\bf 20}, 1599 (1987);
Erratum: {\it J. Phys. A} {\bf 20}, 6159 (1987).

\bibitem{Parisi}
G. Parisi,
 in {\it Proceedings of the Cargr\'ese Summer School} (1973),
unpublished; G.~Parisi,
{\it J. Stat. Phys.} {\bf 23}, 49 (1980).

\bibitem{Prudnikov}
V.V. Prudnikov, P.V. Prudnikov, and A.A. Fedorenko, {\it  J.
 Phys. A} {\bf 32}, L399  (1999);
{\it  J.  Phys. A}  {\bf 32}, 8587 (1999);
{\it Phys. Rev.} {\bf B62}, 8777 (2000).


\bibitem{Hardy48}
G.~H. Hardy {\it Divergent Series} (Oxford, 1948).

\bibitem{borelsummability}
J.-P.~Eckmann, J.~Magnen and R.~S\'en\'eor,
{\it Commun. Math. Phys.}  {\bf 39}, 251 (1975);
J.~S.~Feldman and K.~Osterwalder, {\it Ann. Phys.}  {\bf 97}, 80
(1976); J.~Magnen and R.~S\'en\'eor, {\it Commun. Math. Phys.}
{\bf 56}, 237 (1977); J.-P.~Eckmann and H.~Epstein, {\it Commun. Math.
Phys.}  {\bf 68}, 245 (1979).

\bibitem{Griffiths69}
R. B. Griffiths,
{\it Phys. Rev. Lett.}  {\bf 23}, 17 (1969)

\bibitem{nonsum}
A.~J.~Bray, T. McCarthy, M.~A.~Moore, J.~D.~Reger, and A.~P.~Young,
{\it Phys. Rev.} {\bf B36}, 2212 (1987);
A.~J.~McKane, {\it Phys. Rev.} {\bf B49}, 12003 (1994).

\bibitem{Alvarez00}
G. \'Alvarez, V. Mart\'{\i}n-Mayor, and J. J. Ruiz-Lorenzo,
{\it J. Phys. A} {\bf 33}, 841 (2000).

\bibitem{Lipatov77}
L. N. Lipatov, {\it Zh. Eksp. Teor. Fiz.} {\bf 72}, 411 (1977)
[Sov. Phys. JETP {\bf 45}, 216].

\bibitem{Brezin77}
E. Br\'ezin,  J. Le Guillou, and J.  Zinn-Justin,
{\it Phys. Rev.} {\bf D15}, 1544  (1977).

\bibitem{Brezin78}
E. Br\'ezin and G. Parisi, {\it J. Stat. Phys.} {\bf 19}, 269 (1978).

\bibitem{footnote4}
Note that for {\em finite} number of terms
changing of order of integration and summation can always be performed

\bibitem{Baker}
G. A. Baker, B. G. Nickel, M. S. Green, and D. I. Meiron, {\it
Phys. Rev. Lett.}  {\bf 36}, 1351 (1976); G.~A.~Baker, B.~G.~Nickel,
 and D.~I.~Meiron, {\it Phys. Rev.} {\bf B17}, 1365 (1978).

\bibitem{footnote5}
The construction of the Borel-image for the functions of more
than two variables is performed similarly to the procedure
of the Eq.(\ref{BorIm1}).

\bibitem{Watson74}
P. J. S. Watson,
{\it J. Phys. A}  {\bf 7}, L167 (1974).

\bibitem{Chisholm73}
J. S. R. Chisholm,
{\it Math. Comp.}  {\bf 27}, 841 (1973); G.~A.~Baker~Jn.,
P.~Graves-Morris, {\it  Pad\'e Approximants}, (Addison-Wesley,
Reading, MA, 1981).


\bibitem{Sokolov77}
A. I. Sokolov,
{\it Fiz. Tv. Tela}  {\bf 19}, 748 (1977)
[{\it Sov. Phys.: Solid State}  {\bf 19}, 433].

\bibitem{Sokolov81}
A. I. Sokolov  and B. N. Shalaev, {\it Fiz. Tv. Tela}  {\bf 23}, 2058
(1981) [{\it Sov. Phys.: Solid State}  {\bf 23}, 1200].

\bibitem{Shpot89}
M. A. Shpot,
{\it Phys. Lett. A}  {\bf 142}, 474 (1989).

\bibitem{Mayer89}
I. O. Mayer, A. I. Sokolov,  and B. N. Shalaev, {\it
Ferroelectrics} {\bf 95}, 93 (1989).

\bibitem{Pakhnin00}
D. V. Pakhnin  and A. I. Sokolov, {\it Pis'ma v ZhETF} {\bf 71}, 600
(2000). [{\it JETP Letters}  {\bf 71}, 412]; D.~V.~Pakhnin and
A.~I.~Sokolov, {\it Phys. Rev.} {\bf  B61}, 15130 (2000).

\bibitem{Carmona00}
J. M. Carmona, A. Pelissetto,  and E. Vicari, {\it Phys. Rev.}
{\bf B61}, 15136 (2000).

\bibitem{Holovatch92}
Yu. Holovatch  and M. Shpot, {\it J. Stat. Phys.}  {\bf 66}, 867
(1992).

\bibitem{loopint}
Values of the loop integrals for fixed space dimension $d=2$,
$d=3$ are given in: B.~G.~Nickel, D.~J.~Meiron,  and G.~A.~Baker,
 "Compilation of 2pt and 4pt graphs for continuous spin model",
 University of Guelph Report (1977).

\bibitem{Wilson72}
K. G. Wilson  and M. E. Fisher, {\it Phys. Rev. Lett.} {\bf 28}, 240
(1972).

\bibitem{tHooft72}
G. 't Hooft  and M. Veltman, {\it Nucl. Phys. B}  {\bf 44}, 189
(1972); G.~'t Hooft, {\it Nucl. Phys. B}  {\bf 61}, 455 (1973).

\bibitem{Janssen95}
H. K. Janssen, K. Oerding,  and E. Sengespeick, {\it J. Phys. A}
{\bf 28}, 6073 (1995).

\bibitem{Kleinert95}
H. Kleinert  and V. Schulte-Frohlinde, {\it Phys. Lett. B}
{\bf 342}, 284 (1995).

\bibitem{sqrt}
D. E. Khmel'nitskii, {\it Zh. Eksp. Teor. Fiz.} {\bf 68}, 1960 (1975)
[{\it Sov. Phys. JETP}  {\bf 41}, 981]; A.~B. Harris  and
T.~C.~Lubensky, {\it Phys. Rev. Lett.}  {\bf 33}, 1540 (1974); T. C.
Lubensky, {\it Phys.  Rev.} {\bf B11}, 3573 (1975).

\bibitem{Shalaev97}
B. N. Shalaev, S. A. Antonenko,  and A. I. Sokolov, {\it Phys.
Lett. A}  {\bf 230}, 105 (1997).

\bibitem{Folk99}
R. Folk, Yu. Holovatch,  and T. Yavors'kii, {\it Pis'ma v ZhETF}
{\bf 69}, 698 (1999) [{\it JETP Letters}  {\bf 69}, 747]

\bibitem{Shalaev77}
B. N. Shalaev,
{\it Zh. Eksp. Teor. Fiz.}  {\bf 73}, 2301 (1977)
[{\it Sov. Phys. JETP}  {\bf 46}, 1204].

\bibitem{Jayaprakash77}
C. Jayaprakash  and H. J. Katz, {\it Phys. Rev.} {\bf B16}, 3987
(1977).

\bibitem{Folk98}
R. Folk, Yu. Holovatch, and T. Yavors'kii, {\it J. Phys. Stud.}
{\bf 2}, 213 (1998).

\bibitem{Jug83}
J. Jug ,
{\it Phys. Rev.} {\bf B27}, 609 (1983).

\bibitem{Mayer84}
I. O. Mayer  and A. I. Sokolov, {\it Fiz. Tv. Tela}  {\bf 26}, 3454
(1984) [{\it Sov. Phys.: Solid State} {\bf  26}, 2076]

\bibitem{Mayer89a}
I. O. Mayer,
{\it J. Phys. A}  {\bf 22}, 2815 (1989).

\bibitem{Varnashev00}
K. B. Varnashev,
{\it Phys. Rev.} {\bf B61}, 14660 (2000).

\bibitem{Holovatch98}
Yu. Holovatch  and T. Yavors'kii, {\it J. Stat. Phys.}, {\bf 93}, 785
(1998); {\it Cond. Mat. Phys.} {\bf iss.11}, 87 (1997).

\bibitem{Folk00a}
R. Folk, Yu. Holovatch,  and  T. Yavors'kii, unpublished.

\bibitem{Pelissetto}
A. Pelissetto and E. Vicari, {\it Phys. Rev.} {\bf B62}, 6393
(2000).

\bibitem{errbars}
Error bars of the numerical data obtained by an analysis
of the perturbation theory series result from comparison of
successive orders relative contributions. That is why if
only several initial orders of the expansion are available
the authors usually do not provide error bars  in explicit
form discussing rather the confidence interval (see the upper
part of the Table \ref{table_theory} as well as tables
\ref{chis}, \ref{ramtab1}, \ref{ramtab2}).

\bibitem{Blavats'ka01b}
V. Blavats'ka  and Yu. Holovatch, {\it J. Phys. Stud.}, {\bf 5}, 261
(2001).

\bibitem{Newman82}
K. E. Newman  and E. K. Riedel, {\it Phys. Rev.} {\bf B25}, 264
(1982).

\bibitem{Tissier01} M. T. Tissier, D. Mouhanna, J. Vidal,  and
B. Delamotte, {\it cond-mat/0109176} (2001).

\bibitem{acta}
V. Blavats'ka, C. von Ferber, and Yu. Holovatch, {\it Acta Physica Slovaca} (2002),
to appear.

\bibitem{Dudka01a} M. Dudka, R. Folk,
and Yu. Holovatch,
{\it Condes. Matter Phys.} {\bf 4}, 77 (2001).

\bibitem{noteeps} In order to recover result of Ref. \cite{Aharony75}
one must extract value $\sim 1/\varepsilon$ of one-loop integral from
conventionally normalized couplings.

\bibitem{noteerr}
The second-order $\varepsilon$-expansion contributions to the
fixed point values reported in Refs. \cite{Dudka01a,Dudka01b}
contain some errors. This also causes the deviation of order 5 \%
of the fixed point VIII numerical value for isotropic distribution
\cite{Dudka01a,Dudka}. Here, we give the corrected numbers.

\bibitem{Dudka} M. Dudka, R. Folk,
and Yu. Holovatch, in {\it Fluctuating Path and Fields} ed. by
W.~Janke, A.~Pelster, H.-J.~Schmidt and M.~Bachmann, ( World
Scientific, Singapore, 2001), p.457.

\bibitem{Dudka01b} M. Dudka, R. Folk,
and Yu. Holovatch,  {\it Condens. Matter Phys.} {\bf 4}, 459 (2001).

\bibitem{Mukamel82}
D. Mukamel  and G. Grinstein, {\it Phys. Rev.} {\bf B25}, 381 (1982).

\bibitem{degen} The same reasoning holds also for a formal singularity
at $m=1$ of the fixed point IV of the RAM with isotropic
distribution of the random anisotropy axis.

\bibitem{alpha}
C. Bervillier, {\it  Phys. Rev.} {\bf B34}, 8141 (1986); M. Dudka, Yu.
Holovatch,  and  T. Yavors'kii, {\it J. Phys. Stud.} {\bf 5}, 233 (2001).

\bibitem{Korzhenevskii88}
A. L. Korzhenevskii   and A. A. Luzhkov, {\it Zh. Eksp. Teor.
Fiz.} {\bf 94}, 250 (1988) [{\it Sov. Phys. JETP}  {\bf 67}, 1229].
\end{thebibliography}
\end{document}